\documentclass{vki}

\usepackage{lmodern,pdftexcmds,amsmath}

\usepackage{amsfonts,amsthm,bm} 

\DeclareMathAlphabet{\mathsfbi}{OT1}{\sfdefault}{bx}{sl}
\DeclareMathVersion{sfletters}
\SetSymbolFont{letters}{sfletters}{OML}{ntxsfmi}{b}{it}

\makeatletter
\newcommand{\mathbfsbilow}[1]{%
	\text{\mathversion{sfletters}$\m@th#1$}%
}

\DeclareRobustCommand{\tensor}[1]{%
	\begingroup
	\ifcat\noexpand #1\relax
	\edef\greek@test{\detokenize{#1}}%
	\edef\greek@test{\expandafter\@cdr\greek@test\@nil}%
	\edef\greek@test{\expandafter\@car\greek@test\@nil}%
	\edef\x{\the\lccode\expandafter`\greek@test}%
	\edef\y{\number\expandafter`\greek@test}%
	\ifnum\x=\y\relax
	\mathbfsbilow{#1}%
	\else
	\mathsfbi{#1}%
	\fi
	\else
	\mathsfbi{#1}%
	\fi
	\endgroup
}
\makeatother

\usepackage[most]{tcolorbox}
\usepackage{booktabs}
\usepackage{amsmath}
\usepackage{wrapfig}
\usepackage[linesnumbered,ruled,vlined]{algorithm2e}

\usepackage{mathtools}
\tcbset{enhanced,colback=cyan!2!white,colframe=cyan!90!white,coltitle=blue!20!black,fonttitle=\bfseries}

\makeatother

\usepackage{listings}
\usepackage{hyperref}
\hypersetup{
	colorlinks=true,
	linkcolor=blue,
	filecolor=magenta,      
	citecolor=blue,
	urlcolor=cyan,
}

\usepackage{tikz}
\definecolor{custompurple}{RGB}{102,0,204}
\definecolor{customgreen}{RGB}{0,102,51}
\tikzset{neuron/.style={shape=rectangle, minimum size=1.3cm, inner sep=0.2, draw, font=\normalsize, line width=0.4mm}, 
  input_io/.style={shape=diamond, minimum size=1.7cm, inner sep=0.2, draw, draw=custompurple!90, text=custompurple!90, font=\normalsize, line width=0.4mm}, 
  output_io/.style={shape=diamond, minimum size=1.65cm, inner sep=0.2, draw, draw=customgreen!90,  text=customgreen!90,  font=\normalsize, line width=0.4mm}, 
  dense/.style={shape=ellipse, minimum width=.75cm, minimum height=0.6cm, inner sep=0.2, draw,  font=\normalsize}}
  \tikzset{font=\normalsize}
 \usetikzlibrary{positioning, fit, arrows.meta, shapes}
\begin{document}


\pagenumbering{roman}
\title{\textbf{Prediction of chaotic dynamics from data: An introduction }}
\author{Luca Magri$^{1,2,3}$\thanks{l.magri@imperial.ac.uk}, \; Andrea Novoa$^{1,4}$ \& Elise Özalp$^{1}$ \\ $^{1}$Imperial College London  \\ 
$^{2}$The Alan Turing Institute  \\ 
$^{3}$Politecnico di Torino \\ 
$^{4}$University of Cambridge   
}

\date{\today}
\maketitle

This chapter offers a principled approach to the prediction of chaotic systems from data. First, we introduce some concepts  from  dynamical systems' theory and chaos theory. Second, we introduce machine learning approaches for time-forecasting chaotic dynamics, such as echo state networks and long-short-term memory networks, whilst keeping a dynamical systems' perspective. Third, the lecture contains informal interpretations and pedagogical examples with prototypical chaotic systems (e.g., the Lorenz system), which elucidate the theory. The chapter is complemented by coding tutorials  (online) at \url{https://github.com/MagriLab/Tutorials}. \\

\pagenumbering{arabic}
\setcounter{page}{1}
\clearpage{\pagestyle{empty}} 

\tableofcontents
\clearpage{\pagestyle{empty}} 

\section{Chaotic dynamical systems}  \label{sec:dynamicalsystemsandchaos}
\begin{figure}[h]
    \centering
\includegraphics[width=0.95\textwidth]{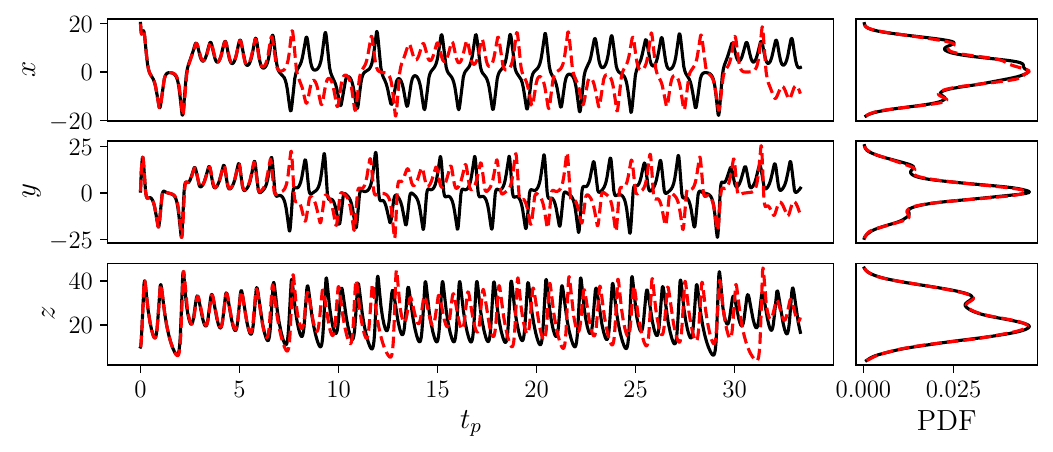}
     \caption{Solution of the Lorenz 63 system solved for $\mathbf{x}_0= [20.0, 1.0, 10.0]$ (black line) and for $\mathbf{x}_0= [20.1, 1.0, 10.0]$ (red dashed line) with a fourth-order Runge-Kutta method. }
    \label{fig:fig1}
\end{figure}

In this lecture, we work with deterministic systems. Deterministic systems are noise-free systems, which means that there exists  one solution that corresponds to an initial condition.  
Chaos is a deterministic phenomenon, which is characterized by erratic behaviour that is difficult---yet possible, in principle---to predict. 
Chaotic dynamics are characterized by the extreme sensitivity to small perturbations, for example, in the initial conditions, or the parameters, or external forcing. 
Two nearby initial conditions, which can differ by a very small amount, will practically diverge in time from each other with an initial exponential rate (Figure~\ref{fig:fig1}). This makes the time accurate prediction of the solution difficult, which is sometimes informally referred to as {\it the butterfly effect}~\citep{lorenz1969atmospheric}. \\

But not all is lost.
The long-term statistics of turbulent flows may be more predictable than the instantaneous time dynamics. The statistics, in fact, may not be significantly affected by tiny perturbations, whereas the instantaneous solution may be. For example, running the same code with the same initial conditions on a different number of processors should, in principle, provide two statistically equivalent solutions\footnote{In this lecture, we work with ergodic systems, see Sec.\ref{sec:att_ergo}.}, but with completely different instantaneous fields after a few time steps (Figure~\ref{fig:fig1}).  \\ 

In this section, we present some basic concepts and nomenclature, which will be used throughout this lecture.  Detailed references in the subject of chaos are \citet{guckenheimer2013nonlinear,hilborn2000chaos,pikovsky2016lyapunov,boffetta2002predictability}, among many others.  

\subsection{Dynamical systems' equations}
We work with chaotic systems that can be described as autonomous dynamical systems as  
\begin{equation}
    \label{eq:system}
		\dot{\mathbf{x}}(t) = \mathbf{F}(\mathbf{x}(t), \mathbf{p}), \quad\quad 
        \mathbf{x}(0) = \mathbf{x}_0
\end{equation}
where the overdot $\dot{(\;)}$ is Newton's notation for time differentiation; 
$\mathbf{x}\in\mathbb{R}^{N_x}$ is the state vector, where the integer $N_x$ denotes the degrees of freedom;
the subscript $0$ denotes the initial condition; 
$\mathbf F: \mathbb{R}^{N_x} \rightarrow \mathbb{R}^{N_x}$ is a 
nonlinear smooth function\footnote{The function ${\mathbf F}$ can represent ordinary differential equations, or partial differential equations that have been spatially discretized  with their boundary conditions.}; 
and $\mathbf{p}$ is a vector containing the system's parameters, which will be dropped unless it is necessary for clarity. %
The solution is given by the trajectory $\mathbf{x}(t)$ that corresponds to the initial condition $\mathbf{x}_0$. Typically, analytical solutions for most problems are not available, which means that we need to resort to discretising time for integration.  \\ 

\subsubsection{Attractors and ergodicity}\label{sec:att_ergo}
We will often talk about ``attractors'' in this lecture. But what is an attractor, $\bar{\mathbf{x}}(t)$? An attractor is the set of values that the solution takes asymptotically. In practice, if you integrate long enough so that the statistics of the solution do not change, you practically obtain (numerically) an approximation of the attractor. Once you are on the attractor, you stay on the attractor (technically, the attractor is an invariant set, which does not change under the dynamical system). 
Chaotic attractors are {\it strange} because they have a zero measure in the embedding phase space and have a {\it fractal} dimension. Trajectories within a strange attractor appear to move around seemingly randomly.   In this lecture, we assume that the system evolves on the same attractor for any initial conditions, for simplicity. In other words, we assume that we work with {\it ergodic systems}~\citep{birkhoff1931}, in which the initial condition does not influence the attractor, therefore, the time average is equal to the ensemble average. 

\subsection{Linear analysis}\label{sec:lin_analys_sec}
We can say, informally, that if you sufficiently zoom in a plot of a nonlinear function, you will find a straight line. Likewise, we can say that if you sufficiently zoom in a nonlinear dynamical system, you will find a linear behaviour.  This means that a nonlinear attractor can be ``tessellated'' or ``patched'' by infinitesimally small straight planes. Dynamically, we can say that the journey of any state starts from the tangent space (i.e., the small straight patches), which is why it is a key element to characterize. 

In the tangent space, the evolution of a state is, of course, given by the dynamical system, $\dot{\mathbf{x}}=\mathbf{F}(\mathbf{x})$. However, {\it in the limit of infinitesimal perturbations},  the dynamical system is identical to a linear system $\dot{\mathbf{x}}=\mathbf{F}(\mathbf{x}) =\mathbf{J}\mathbf{x}$ ($\mathbf{J}$ is the Jacobian), which greatly simplifies our analysis. 
{\it
The properties of the tangent (linear) space determine many properties of the nonlinear solution.} This is the objective of stability analysis. 
\\

In {\it stability analysis}, we are interested in computing the evolution of infinitesimal perturbations around a reference point of the attractor, $\bar{\mathbf{x}}(t)$. 
To do so, we split the solution as  
\begin{align}\label{eq:split_sol}
\mathbf{x}(t) = \bar{\mathbf{x}}(t) + \mathbf{x}'(t), 
\end{align}
where $\bar{\mathbf{x}}(t)$ is the unperturbed solution\footnote{Note that we have made no assumption on the unperturbed solution that lies on the attractor, i.e., it may be time-dependent and aperiodic.} of~\eqref{eq:system} such that $\lvert\lvert{\mathbf{x}}'(t)\rvert\rvert\sim O(\epsilon)$, where $\epsilon\rightarrow0$. 
The perturbation equation is found by truncating the Taylor expansion of the dynamical equations~\eqref{eq:system} to the first order, which yields the tangent equation
\begin{align}\label{eq:tangent_eq}
\dot{\mathbf{x}}' =  \mathbf{J}(t)\mathbf{x}', \quad\quad \mathbf{x}'(0) = \mathbf{x}'_0. 
\end{align}
where $\mathbf{J}(t)\equiv \frac{d\mathbf{F}}{d\mathbf{x}}\big|_{\bar{\mathbf{x}}(t)}$ 
is the Jacobian\footnote{In tensor notation, $J_{ij} \equiv dF_{ij}/{dx_j}\big|_{\bar{\mathbf{x}}(t)}$, with  $i,j=1,2,\ldots,N_x$.}. 
The Jacobian is a key quantity in dynamical systems. On the one hand, the Jacobian around a fixed point is constant (a fixed point could be a steady solution of Navier-Stokes equations). The eigenvalues and eigenvectors of the Jacobian will establish the stability behaviour. 
If at least one eigenvalue has a positive growth rate, the fixed point is linearly unstable. 
Around a periodic flow, the Jacobian matrix is periodic. 
(A periodic flow could be a periodic solution of Navier-Stokes equations.)
If at least one eigenvalue (Floquet exponent) has a positive growth rate, the periodic solution is linearly unstable~\citep[e.g., for linear flow analysis, ][]{magri2019adjoint,magri2023linear}. On the other hand, what happens when we perform linear analysis on chaotic solutions? The Jacobian is chaotic, thus, we need to generalize stability analysis to chaotic Jacobians. 

\subsection{Largest (dominant) Lyapunov exponent}\label{subsec:largest_lyap_exponent}
We analyse what makes a  solution chaotic.
We analyse how a small\footnote{We will use the word ``small'' to informally mean ``infinitesimal'', unless otherwise specified. } perturbation evolves. 
For this, it is convenient to introduce the tangent propagator, which formally maps the perturbation, $\mathbf{x}'$, from time $t$ to time $\tilde{t}$, as
\begin{align}
    \mathbf{x}'(t+\tilde{t}) = \mathbf{M}(t,\tilde{t}) \mathbf{x}'(t).   
\end{align}
\begin{tcolorbox}[breakable, opacityframe=.1, title=Equation of tangent propagator]
Show that 
\begin{equation}
\begin{cases}
	\frac{d\mathbf{M}}{d\tilde{t}} = \mathbf{J}(\tilde{t}) \mathbf{M}, \\
	\mathbf{M}(t,0) = \mathbf{I}, 
	\end{cases}
    \label{eq:tangent_system}
\end{equation}
where $\mathbf{I}$ is the identity matrix. Show that   \begin{align}
    \mathbf{M}(t, \tilde{t})& = \mathcal{P}\left(\exp\left({\int_t^{t+\tilde{t}} \mathbf J(\chi) d\chi}\right)\right), 
\end{align}
where $\mathcal{P}$ is the path-ordering operator. 
\end{tcolorbox}
%
%
Setting $t=0$ without loss of generality, the norm of an infinitesimal perturbation, $\mathbf{x}_0'$, applied to the unperturbed solution, $\bar{\mathbf{x}}_0$, asymptotically grows as~\citep{oseledets1968multiplicative}  
\begin{equation}
	||\mathbf{x}'(\tilde{t})|| \cong ||\mathbf x_0'|| e^{\lambda_1(\mathbf{x}'_0 , \bar{\mathbf x}_0) \tilde{t}},
\end{equation}
where $\cong$ means ``asymptotically equal to''. This is a result of Oseledets’ theorem~\citep{oseledets1968multiplicative}. For practical purposes, we can think of chaotic systems as those that have (at least one) positive Lyapunov exponent\footnote{There are pathological examples of systems that do not fulfill this criterion. In practice, for engineering systems, it is safe to say that at least one positive Lyapunov exponent is associated with chaos, and vice versa.}.
Furthermore, Oseledets’ theorem  shows that the Lyapunov exponents are constants of the attractor, and, in ergodic systems, they do not depend on the initial condition, $\mathbf{x}'_0$. Therefore
\begin{equation}
	\label{eq:lyapunov_exponent}
	\lambda_1 = \lim_{\tilde{t} \rightarrow \infty} \frac{1}{\tilde{t}} \log \frac{||\mathbf{M}(0, \tilde{t}) \mathbf{x}'_0 ||}{||\mathbf{x}'_0 ||} 
\end{equation}
is the largest (dominant) Lyapunov exponent, which is the time-averaged growth rate of infinitesimal perturbations. 
The largest Lyapunov exponent indicates the type of solution (i.e., attractor). 
If the largest Lyapunov exponent is $\lambda_1<0$,  perturbations decay and the attractor is a fixed point. 
If $\lambda_1 = 0$, the attractor is a periodic orbit. If $\lambda_1 = 0 $, the perturbation grows exponentially and, typically, the attractor is chaotic. These criteria can be used to classify bifurcations in fluid systems~\citep{Huhn2020}. The largest Lyapunov exponent is a practical measure to compute the predictability of large-scale simulations because it (i) is easy to calculate, and (ii)  does not depend on the initial conditions in ergodic processes. In large-scale fluid-dynamics simulations, the Lyapunov exponent was calculated in channel and bluff-body flows~\citep{Blonigan2016}, homogeneous isotropic turbulence~\citep{Nastac2017,mohan2017scaling}, reacting and non-reacting turbulent jets~\citep{Nastac2017}, a two-dimensional airfoil~\citep{fernandez2017lyapunov}, backward-facing step~\citep{Ni2017}, partially-premixed flames~\citep{Hassanaly2019}, to name only a few. %
\subsubsection{Practical computation of the dominant Lyapunov exponent}\label{subsec:algo}
\begin{figure}[!htb]
    \centering
    \includegraphics[width=0.8\textwidth]{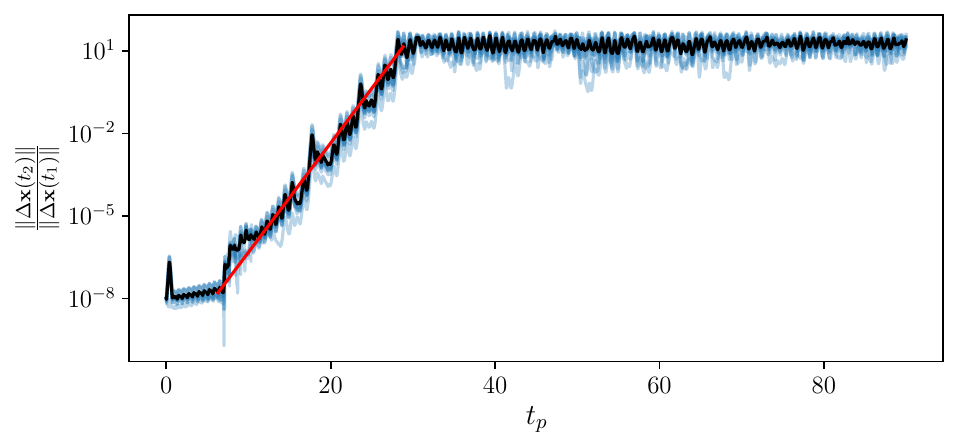}
    \caption{Separation trajectory in the Lorenz system over $10$ random perturbations (blue lines) and their mean (black line). The dominant Lyapunov exponent is $\lambda\approx0.929$, which corresponds to the slope (red line).}
    \label{fig:lorenz:lyapunov:growth}
\end{figure}

  We focus on the dominant Lyapunov exponent, $\lambda_1$. 
    Obtaining accurate estimates of the Lyapunov exponent is straightforward even in large-scale simulations. A non-intrusive method is based on the calculation of the \textit{separation} trajectory, also known as the \textit{error} trajectory. 
    The separation trajectory is the difference between two nearby trajectories (which can be Eulerian fields in computational fluid dynamics), which originate from two close initial conditions.  Because it is almost sure\footnote{``Almost sure'' means that the Lebesgue measure of the set is 1, i.e., the probability of ``being sure'' is 1.} for the separation trajectory to have a component--even minuscule--in the direction that will grow with the dominant Lyapunov exponent, the separation trajectory will almost surely grow at an exponential divergence rate provided by the dominant Lyapunov exponent.  This is why the dominant Lyapunov exponent is of paramount importance in chaotic flows. It can be calculated as described in the following practical and non-intrusive algorithm. 
\begin{tcolorbox}[breakable, opacityframe=.1, title=Pseudoalgorithm for dominant Lyapunov exponent ]
    \begin{enumerate}
        \item \textbf{Statistically converged solution.} Run a numerical simulation~\eqref{eq:system} until statistical convergence is reached, say, at $t_0$. The time solution thereafter  approximates the attractor, $\bar{\mathbf{x}}(t)$.  
        \item \textbf{Reset time}, $t=t_0$.
        \item \textbf{Perturb}. At $t=t_0$, impose the perturbed solution $\mathbf{x}'=\boldsymbol{\epsilon}$ as 
            \begin{equation} 
    \label{Perturbation}
    \mathbf{x}'(t_0) = \bar{\mathbf{x}}(t_0) + \boldsymbol{\epsilon}, 
    \end{equation}
    where $\boldsymbol{\epsilon}$ is a small random field, whose norm is typically in the range $10^{-9}-10^{-3}$ . 
        \item \textbf{Separation trajectory.} Advance both solutions, $\bar{\mathbf{x}}(t_0)$ and $\mathbf{x}'(t_0)$, to some time $t_f$ and evaluate the separation trajectory 
            \begin{equation}\label{eq:deltaphiddd}
    \Delta \mathbf{x}(t)= \mathbf{x}' (t)-\bar{\mathbf{x}}(t)  \quad\quad t_0\leq t\leq t_f.
    \end{equation} \label{item:tf}
    \item \textbf{Identification of the linear region} $t_1\leq t \leq t_2$ where $\log \left( \Vert\Delta\mathbf{x}(t)\Vert \right)$ grows linearly.     $t_f$ in item~\ref{item:tf} must be larger than $t_2$.    
    %
    %
        \item \textbf{Lyapunov exponent.}  The Lyapunov exponent is the slope of the linear region, which can be obtained by linear regression 
        \begin{align}\label{eq:llkkkjja1}
        \lambda_1 \approx \frac{1}{t_2-t_1}\log{\left(\frac{\Vert\Delta\mathbf{x}(t_2)\Vert}{\Vert\Delta\mathbf{x}(t_1)\Vert}\right)}.
        \end{align} 
    \end{enumerate}

    \textbf{Practical tip:} The evolution of the error trajectory can be quite erratic (loosely put, it might look ``noisy'', see Figure~\ref{fig:lorenz:lyapunov:growth}). In this case, you can run an ensemble of computations and take the average to estimate the Lyapunov exponent. 
    \end{tcolorbox}

\subsection{Lyapunov spectrum}
So far we have defined the largest Lyapunov exponents. What about the ``other'' Lyapunov exponents in the spectrum? Oseledets theorem shows that, for $\mathbf{x}\in\mathbb{R}^{N}$ and non-degenerate systems, there exist $N-$Lyapunov exponents, $\lambda_1 \geq \dots \geq \lambda_N$. The Lyapunov spectrum is provided, theoretically, by the eigenvalues of the Oseledets matrix~\citep{oseledets1968multiplicative} 
\begin{equation}
   	\mathbf{\Xi^{\pm}}(t) = \lim_{t' \rightarrow \pm \infty} \frac{1}{2t'} \log \left[ \mathbf{M}(t,t')^\mathrm{T}  \mathbf{M}(t,t') \right]. 
	\label{eq:oseledets_operator}
\end{equation}
This matrix is called ``forward'' if $t' \rightarrow +\infty$ or ``backward'' if $t' \rightarrow -\infty$. 
%
Insight into the Oseledets' matrix can be obtained by applying a singular value decomposition to $\mathbf{M}$, $\mathbf{M}(t, t') = \mathbf U \mathbf{S} \mathbf{V}^\mathrm{T}$, where $\mathbf U$ and $\mathbf{V}$ are orthogonal matrices and $\mathbf{S}$ is a diagonal matrix with non-negative real entries (the singular values).
Substituting the decomposition in the argument of the logarithm of \eqref{eq:oseledets_operator}, an eigenvalue decomposition is obtained,
$\mathbf{M}^\mathrm{T} \mathbf{M} = \mathbf{V} (\mathbf{S}^\mathrm{T} \mathbf{S}) \mathbf{V}^\mathrm{T} = \mathbf{V} \mathbf{S}^2 \mathbf{V}^\mathrm{T}$, which, after applying the logarithm, becomes $\mathbf{V} \log(\mathbf{S}^2) \mathbf{V}^\mathrm{T} = 2 \mathbf{V} \log(\mathbf{S}) \mathbf{V}^\mathrm{T}$. Thus, \eqref{eq:oseledets_operator} can be rewritten as
\begin{equation}\label{eq:sin_oseld_ein}
	\mathbf{\Xi}^\pm(t) = \lim_{t' \rightarrow \pm \infty} \mathbf{V} \frac{ \log\left[\mathbf{S}(t,t')\right] }{t'} \mathbf{V}^\mathrm{T},
\end{equation}
which shows that the Lyapunov exponents, which are the eigenvalues of $\mathbf{\Xi}^\pm$, are equal to the average logarithms of the singular values of $\mathbf{M}(t, t')$.
However, the numerical computation of the Lyapunov spectrum from~\eqref{eq:sin_oseld_ein} is unstable: You will almost never get a good prediction of the Lyapunov spectrum~\citep[for a discussion refer to][]{Huhn2020,huhn2022optimisation}. This is because we need a long-term averaging (in principle a limit to infinity) to obtain the Lyapunov spectrum. Because of this long integration, vectors almost surely align to the dominant direction corresponding to the largest Lyapunov exponent, $\lambda_1$. To overcome this numerical overflow, we employ the Gram-Schmidt orthonormalisation~\citep{schmidt1907theorie,bennetin1980lyapunov,sandri1996numerical}.  

\subsubsection{Practical computation of the Lyapunov spectrum}\label{sec:practical_computation_le}
We know how to define and compute the largest Lyapunov exponents, from~\eqref{eq:lyapunov_exponent} and~\eqref{eq:llkkkjja1}. However, we know that there exist $N$ Lyapunov exponents in chaotic systems, under mild assumptions. We need first to define these. Let us imagine a Kafkaesque situation: 
You are sitting on a chaotic attractor at point, $\bar{\mathbf{x}}_0$. You and your friends are infinitesimal, therefore, you can only walk on straight paths. Now you ask your friends to surround yourself, and you arrange everybody to fill a perfect parallelepiped centred at  $\bar{\mathbf{x}}_0$. Now, the dynamical system, $\mathbf{F}$ (or, equivalently, $\mathbf{J}$ because we are infinitesimal) will take you from $\bar{\mathbf{x}}_0$ to a close location $\bar{\mathbf{x}}_1$. In this trip, some of you will stretch at different rates, some others will not change, and others will be compressed at different rates.
Let us start with the lucky ones: Those who do not change. These are you and the people close to you in the direction of motion $\dot{\mathbf{x}}=\mathbf{F}(\mathbf{x})$. 
Let us move to the unlucky ones, i.e., those who are stretched or compressed by $\mathbf{F}$ (or, equivalently, $\mathbf{J}$ because we are infinitesimal). Some will get stretched (or compressed) more than others: It depends in which directions they sit around you. 
So, let us formalize this absurd example with some more rigour.
\\ 

\begin{figure}
    \centering
        \includegraphics[width=\textwidth]{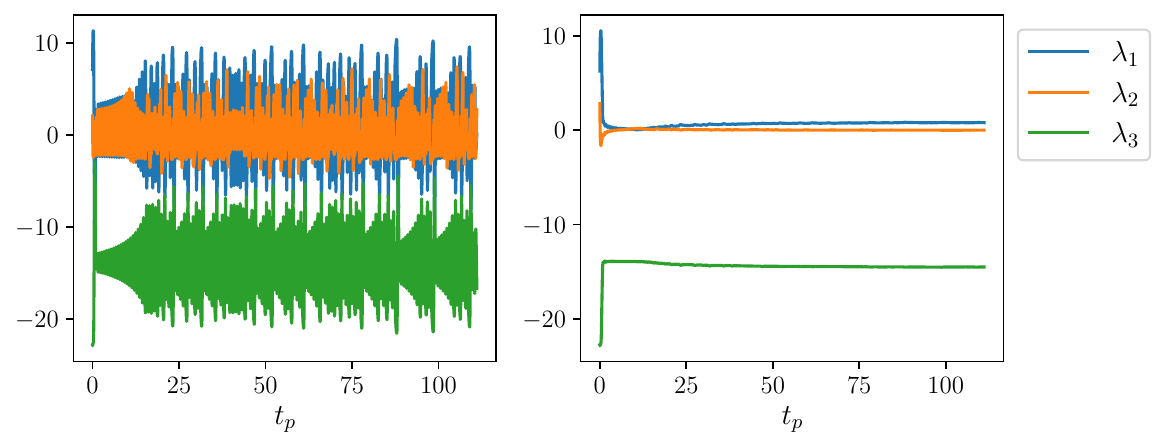}
\caption{Instantaneous Lyapunov exponents (left) and moving average of the Lyapunov exponents (right) over 100 $t_p$ for the Lorenz 63 system. The time $t_p$ is the physical time normalized by the dominant Lyapunov exponent (Sec.~\ref{sec:lyap_exp})}
    \label{fig:lyapunov_exponents_convergence}
\end{figure}

First, let us consider an infinitesimal $p$-volume centred {\it in the tangent space}. A $p$-volume is just a parallelepiped that might have as many sides, $\mathbf{t}_i$, as the phase space, or less, hence, $i=1,2,\ldots,p\leq n$
\begin{align}
\mathrm{Vol}^{(p)}(\mathbf{t}_1, \mathbf{t}_2, \ldots, \mathbf{t}_p) \equiv \mathbf{t}_1 \wedge \mathbf{t}_2 \wedge \ldots \wedge \mathbf{t}_p, 
\end{align}
where $\wedge$ is the wedge symbol (which is the cross product operation in a three-dimensional space). 
Second, on average, the volume will expand with a rate given by the average divergence, $\lambda^{(p)}$ 
\begin{align}\label{eq:volp_ly}
\lambda^{(p)} \equiv \lim_{t\rightarrow \infty}\frac{1}{t}  \log\left[ \frac{\mathrm{Vol}^{(p)}(\mathbf{M}(t)\mathbf{t}_1, \mathbf{M}(t)\mathbf{t}_2, \ldots, \mathbf{M}(t)\mathbf{t}_p)}{\mathrm{Vol}^{(p)}(\mathbf{t}_1, \mathbf{t}_2, \ldots, \mathbf{t}_p)} \right]. 
\end{align}
Because the initial volume is arbitrary, we take it to be unitary. 
``Expansion'' becomes  ``compression'' if the divergence sign is negative. 
Third, we observe that the vectors $\mathbf{M}\mathbf{t}_i$ are, in general, non-orthogonal, therefore, they form a parallelepiped. This parallelepiped will numerically collapse along the direction associated to the dominant Lyapunov exponent, which is what we want to avoid to compute the Lyapunov spectrum. However, we know something about geometry: {\it We know that the volume of a parallelepiped is equal to the volume of the equivalent rectangular parallelepiped.} Therefore, we will compute the volume divergence of the equivalent rectangular parallelepiped, and this will be it. Fourth, how do we make an oblique parallelepiped rectangular? From a linear algebra point of view, this question can be phrased as ``Given a non-orthogonal basis, how do we make it orthogonal?''. The answer lies in the Gram-Schmidt orthonormalisation.
So we rectangularize\footnote{In all likelihood, this is a neologism.} the initial volume with Gram-Schmidt procedure
\begin{align}
(\mathbf{q}_1(t),\mathbf{q}_2(t), \ldots, \mathbf{q}_p(t), ) \leftarrow (\mathbf{t}_1(t),\mathbf{t}_2(t), \ldots, \mathbf{t}_p(t)),
\end{align}
where $\mathbf{q}_i$ are orthogonal to each other. This operation preserves the volume 
\begin{align}
\mathrm{Vol}^{(p)}(\mathbf{M}(t)\mathbf{t}_1, \mathbf{M}(t)\mathbf{t}_2, \ldots, \mathbf{M}(t)\mathbf{t}_p) = \mathrm{Vol}^{(p)}(\mathbf{q}_1(t), \mathbf{q}_2(t), \ldots, \mathbf{q}_p(t)).
\end{align}
Because the vectors $\mathbf{q}_i$ span a rectangular parallelepiped, the computation of the volume is straightforward
\begin{align}
\mathrm{Vol}^{(p)}(\mathbf{q}_1(t), \mathbf{q}_2(t), \ldots, \mathbf{q}_p(t)) = || \mathbf{q}_1(t)|| || \mathbf{q}_2(t)|| \ldots || \mathbf{q}_p(t)|| 
\end{align}
Hence, we can estimate the average divergence by averaging over a sufficient number $S$ of short time windows $T$, {in which we perform repeated orthonormalisation at the beginning of each window}
\begin{align}
\lambda^{(p)}\approx \frac{1}{kT}\sum_{k=1}^{S}  \log\left[ || \mathbf{q}_1(t)|| || \mathbf{q}_2(t)|| \ldots || \mathbf{q}_p(t)|| \right]
\end{align}
The $i$-th Lyapunov exponent is, thus, the average stretching rate of the $i$-th side of the parallelepiped 
\begin{align}
\lambda_i\approx \frac{1}{kT}\sum_{k=1}^{S} \log\left[ || \mathbf{q}_i(t)||  \right]
\end{align}
Gram-Schmidt orthonormalisation is practically encapsulated in the QR algorithm. Therefore, to compute the Lyapunov spectrum, we perform QR decomposition as shown in the pseudoalgorithm below. The Lyapunov exponents of the Lorenz system are shown in Table~\ref{tab:lyapunov_lorenz} and Figure~\ref{fig:lyapunov_exponents_convergence}. 

\begin{table}[!h]
\begin{center}\label{tab:lyapunov_lorenz}
\begin{tabular}{|p{3cm}|p{3cm}|p{3cm}|}
 \hline
 \multicolumn{3}{|c|}{Lorenz 63} \\
 \hline
   $\lambda_1$ & $\lambda_2$ & $\lambda_3$ \\
 \hline
   $0.9050$ &  $9 \times 10^{-5}$ & $-14.572$   \\
 \hline
\end{tabular}
\end{center}
\caption{\label{le-table}Lyapunov of the Lorenz 63 system.}
\end{table}

 \begin{tcolorbox}[breakable, opacityframe=.1, title=Algorithm: Computing Lyapunov spectrum with Gram-Schmidt orthomalisation.
 ]

\textbf{Initialisation:}
\begin{enumerate}
    \item {Initialize $N$ Gram-Schmidt vectors:} $\mathbf{U} \gets \textit{random}\in \mathbb{R}^{N_x\times N_x}$ 
     \item {Orthonormalize Gram-Schmidt vectors:} $\mathbf{Q}, \mathbf{R} \gets QR(\mathbf{U})$
     \item {Update Gram-Schmidt vectors:} $\mathbf{U} \gets \mathbf{Q}$
\end{enumerate}

\noindent \textbf{Evolve the solution and GSV simultaneously for $N^{lyap}$ steps. Discard a transient.}
\begin{enumerate}
    \item {Evolve the system, Eq.~\eqref{eq:system} } $ \mathbf{x}(t_{i+1}) = \textrm{Integrate}( \mathbf{F}, \mathbf{x}(t_{i})) $
     \item {Evolve the tangent propagator:} $ \mathbf{U} \gets  \mathbf{M} \mathbf{U}$
     \item {Orthonormalize and update Gram-Schmidt vectors:} $ \mathbf{Q},  \mathbf{R} \gets QR( \mathbf{U});  \mathbf{U} \gets  \mathbf{Q}$
     \item {Track Lyapunov exponents:} $ \mathbf{\lambda}[:, i] \gets \log(\textrm{diag}( \mathbf{R}[i, i]))/\Delta t$ 
\end{enumerate}
\textbf{Time-averaged Lyapunov exponents: } $ \mathbf{\lambda}_j \gets \sum_{i=0}^{N_{QR}}  \mathbf{\lambda}[j, i]/{(N^{lyap}\Delta t)}$ 
\end{tcolorbox}

 \subsection{Metrics and indicators of chaos} 

Dynamical systems theory provides the predictability of a chaotic simulation, which is the {\it average} time scale after which the trajectories diverge due to the butterfly effect. 
There exist different approaches to characterize a chaotic solution~\citep{ruelle1979ergodic,Eckmann1985,boffetta2002predictability}. On the one hand, geometric approaches estimate the dimension of the chaotic attractor, which provides an estimate of the active degrees of freedom of the chaotic dynamical system. An accurate measure is the Hausdorff dimension \citep{farmer1983dimension}, which is often approximated by box counting, or by an upper bound with the Kaplan-Yorke dimension~\citep{frederickson1983liapunov}.  
           On the other hand, dynamical approaches estimate the entropy content of the solution, for example via the Kolmogorov-Sinai entropy, and the separation rate of two nearby solutions via the Lyapunov exponents~\citep{boffetta2002predictability}.  In this lecture, we  describe dynamical systems' concepts, which can be used as metrics to practically assess machine learning algorithms for chaotic time series forecasting. We focus on Lyapunov exponents to evaluate both the dynamical content and the geometric dimension of the attractor.  Lyapunov exponents play a central role because they underpin a variety of chaotic properties.  

\subsubsection{Lyapunov time} \label{sec:lyap_exp}
 The {\it predictability} is  a key time scale of chaotic dynamical systems, which can be defined as the \textit{Lyapunov time}, which in turn, is defined as the inverse of the Lyapunov exponent 
        \begin{align}
        t_p \equiv \frac{1}{\lambda_1}. 
        \end{align}
The Lyapunov time is a {\it scale}, which, 
         from Eq.~\eqref{eq:llkkkjja1}, is the average time that the separation trajectory's norm takes to get amplified by $e\approx2.718$. Physically, the predictability provides a time scale for the divergence of two nearby trajectory due to the chaotic nature of turbulent flows\footnote{The Lyapunov exponent was proposed as a metric to assess the quality of a large-eddy simulation~\cite{Nastac2017}: If grid $a$ has the same Lyapunov exponent as grid $b$, the grid with fewer degrees of freedom can be used to predict chaotic behaviour.}.  

\subsubsection{Kaplan-Yorke dimension} 
We wish to have a metric that captures the attractor's dimension. 
Chaotic attractors have a fractal structure  and their dimensions can be estimated through the Lyapunov exponents. (The explanation of fractal sets is beyond the scope of this lecture. Please, just bear in mind that the dimension of a fractal set is not an integer.) The Kaplan-Yorke conjecture proposes an estimate (upper bound) of the attractor's dimension as~\citep{frederickson1983liapunov, kantz2004nonlinear} 
\begin{equation}
    D_{KY} = k + \frac{\sum_{i=1}^k \lambda_i}{|\lambda_{k+1}|}
\end{equation}
 with $\sum_{i=1}^k \lambda_i > 0 $ and $\sum_{i=1}^{k+1} \lambda_i < 0 $. 
This relationship relates dynamics (Lyapunov exponents) to the attractor's geometry. While a proof of the conjecture is not available for general cases, the Kaplan-Yorke (K-Y) dimension is {\it de-facto} the practical way of estimating the dimension of a strange attractor when you can compute the Lyapunov spectrum (or the portion that is necessary for the K-Y dimension). If you cannot compute the Lyapunov spectrum, you can use the correlation dimension~\citep{hilborn2000chaos} to estimate the attractor's dimension. 

\subsubsection{Lyapunov spectrum} 
The sum of all Lyapunov exponents (LEs) measures the expansion rate of volumes in the whole phase space, i.e., the divergence (Sec.~\ref{sec:practical_computation_le}). In dissipative systems the sum of the Lyapunov exponents is negative, which means that volumes visited by generic trajectories shrink exponentially to zero. Therefore, trajectories converge to an attractor.  On the other hand, in conservative systems, the sum of the Lyapunov exponents~\eqref{eq:volp_ly} is zero, i.e., volumes are preserved (also known as Liouville theorem). 

\begin{tcolorbox}[breakable, opacityframe=.1, title=Limitations of Lyapunov exponents]
 Lyapunov exponents capture asymptotic and average behaviour of the chaotic solution. First, there exist finite-time fluctuations, which are important for the characterization of local predictability (i.e., some areas of the attractors might be more predictable than others). The generalized Lyapunov exponents have been introduced with the purpose of taking this into account. Second, the Lyapunov exponents are defined in terms of infinitesimally close trajectories (i.e., linear dynamics). Extension to finite-amplitude perturbations can be found in~\citep{boffetta2002predictability} shows both the local Lyapunov exponents and the averaged quantities.  
 \end{tcolorbox}
 
 \subsubsection{Statistics}\label{sec:chaos_stats}
\begin{figure}[!ht]
    \centering
    \includegraphics[width=0.9\textwidth]{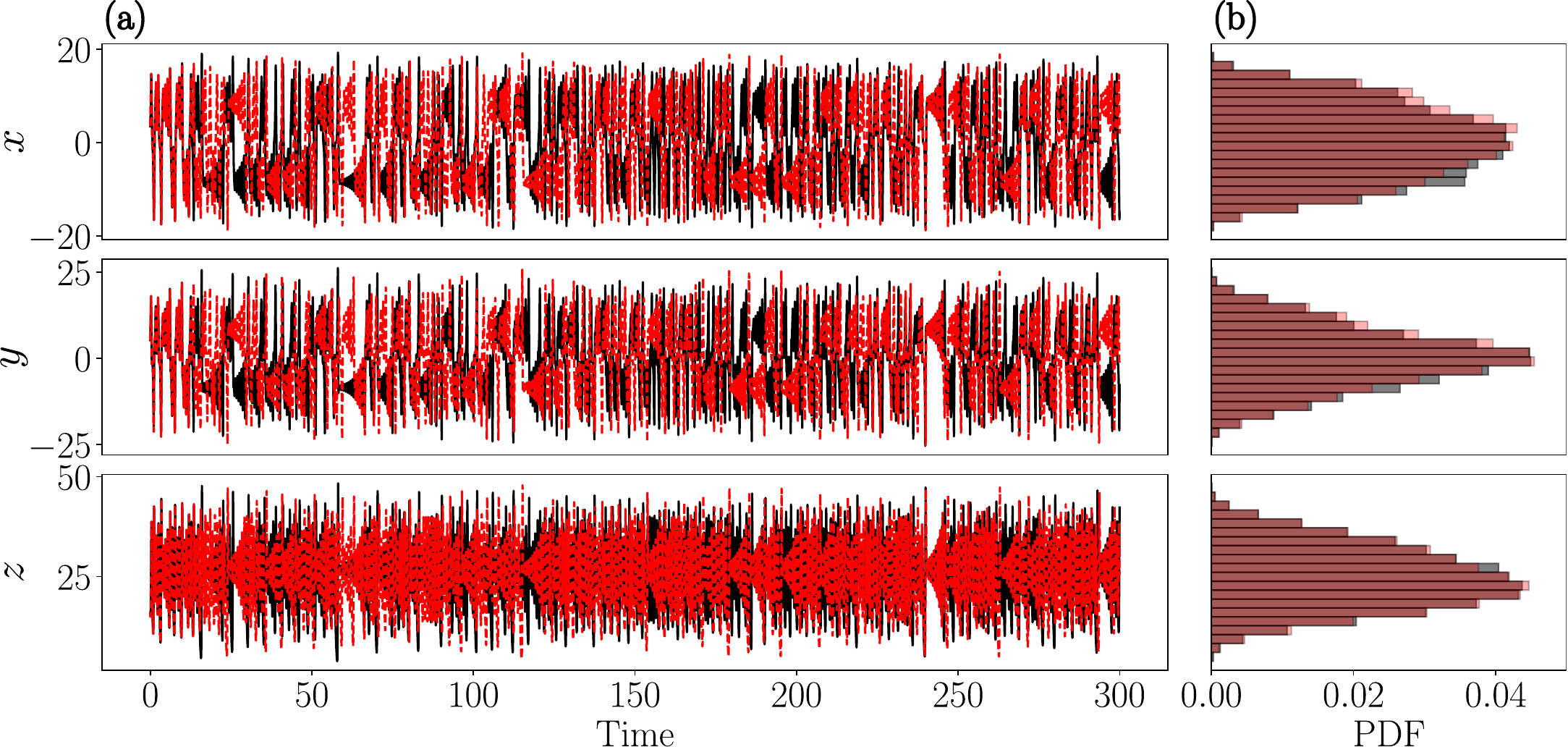}
    \caption{(a) Two long time series for the Lorenz system. (b) Probability density function of the state's components~\citep{racca2023neural}.}  
\label{fig:chaos_stats}
\end{figure}

Other ways of characterising chaotic solutions are the statistics   of the signal. 
In ergodic systems,
the statistics are not affected by the butterfly effect \citep{eckmann1985ergodic}. Every solution of the system (trajectory) eventually visits all parts of the attractor, and different trajectories share the same long-term (infinite-time) statistics (Fig. \ref{fig:chaos_stats}). This means that, although the time-accurate prediction of chaotic dynamics is sensitive to initial conditions, the statistical prediction is not. Therefore, chaotic dynamics can be  described by their statistics, which can be accurately computed from a single trajectory of the system that lasts for sufficiently long times. Mathematically, this is expressed by an equality between the expected value, $\mathbb{E}\left[k \right]$, of the observable, $k(\mathbf{x}(t))$, and its time average over the trajectory, $\mathbf{x}(t')$, (Birkhoff ergodic theorem  \citep{birkhoff1931})
\begin{equation}
    \mathbb{E}\left[k\right] = \lim_{t\to\infty}\frac{1}{t}
    \int_0^\mathrm{T}k(\mathbf{x}(t'))\mathrm{d}t'.
\end{equation}
 
Finally, there are other good indicators of chaos, e.g., the Kolmogorov-Sinai~\citep{boffetta2002predictability}. Most of these additional metrics can be estimated from the knowledge of the Lyapunov exponents.
%


\section{Machine learning for dynamical systems}
We offer a principled approach to introduce machine learning methods for time series forecasting (RNNs, LSTMs, ESNs).   
We will first analyse the formal solution of a dynamical system, thereby justifying the choice of sequential machine learning approaches. Whether we use RNNs, LSTMs, or ESNs, or else, we need to keep in mind that the objective is to develop a data-driven method that is an accurate approximation of a dynamical system given data. 
So, a good starting point is to  start from what we know, which is the dynamical equation~\eqref{eq:system}. 
%
%
When we consider the dynamical system at discretized time instants, $t_i=i\Delta t$, with $i=0,1, 2,\ldots,(N-1)$, the analytical solution is 
\begin{align}\label{eq:ML_24}
\mathbf{x}(t_{i+1}) = \mathbf{x}(t_i) + \int_{t_i}^{{t_{i+1}}}\mathbf{F}\left(\mathbf{x}(t)\right)dt.
\end{align}
With no approximation being made, we can expand~\eqref{eq:ML_24} with a Taylor expansion
\begin{align}\label{eq:ML_25}
\mathbf{x}(t_{i+1}) = 
\mathbf{x}(t_i) + 
\mathbf{F}\left(\mathbf{x}(t_i)\right)
\Delta t + \mathcal{O}\left(\Delta t^2\right).
\end{align}
Let us now consider a sufficiently small $\Delta t$, so we can ignore the negligible terms $\mathcal{O}\left(\Delta t^2\right)$. We can recast Eq.~\eqref{eq:ML_25}, which is amenable to interpretation 
\begin{align}\label{eq:ML_26}
\mathbf{x}(t_{i+1}) = 
\mathbf{x}(t_{i}) + 
\mathbf{F}\Big(
\mathbf{x}(t_{i-1})
+\mathbf{F}\left(\mathbf{x}(t_{i-1})\right)\Delta t
\Big)
\Delta t.
\end{align}
The formal solution tells us that
(i) the future ($\mathbf{x}(t_{i+1})$) is equal to the present ($\mathbf{x}(t_i)$) plus a correction $\mathbf{F}\left(\mathbf{x}(t_i)\right)
\Delta t$; 
(ii) the correction depends on the past (there is memory);  
(iii) the dependence on the past is recursive; and 
(iv) the dependence on the past is nonlinear. These observations partly inspire the design of machine learning methods that are suitable for dynamical systems. \\

We briefly introduce Recurrent Neural Networks (RNNs) in Sec.~\ref{sec:rnns}; and explain in more depth Echo State Networks (ESNs) in Sec.~\ref{sec:esns}, and Long Short-Term Memory (LSTM) networks in Sec.~\ref{sec:lstms}, with their physics-constrained architectures. These neural networks are designed so that they retain a {\it memory} of the inputs to imitate the behaviour of dynamical systems where the evolution depends on the history of the state.

\section{Recurrent Neural Networks}\label{sec:rnns}
In time series prediction, the data is sequentially ordered in time. 
In an RNN, similarly to feedforward neural network, neurons (or units) are connected through links which enable activations to propagate through the network. However, in contrast to feedforward neural networks, the connection within RNN have cycles meaning that the neurons contain a feedback loop. The existence of these cycles enables the RNN to develop a self-sustained temporal activation dynamics (hence RNNs are dynamical systems) and to possess a dynamical memory of the input excitation.
Because of the long-lasting time dependencies of the internal state, however, training RNNs with backpropagation through time is notoriously difficult \citep{werbos1990backpropagation}: The gradient either vanishes or explodes. To circumvent the gradient instability,  echo state networks and long-short term memory networks were introduced. 

%
%


\section{Echo state networks}\label{sec:esns}
 
Echo State Networks (ESN) are a form of reservoir computing\footnote{ The basis of reservoir computing was introduced and developed independently by Wolfgang Mass \citep{Maass2001} with liquid state machines, and Herbert Jaeger~\citep{Jaeger2001} with Echo State Networks.} The approach is motivated mainly by a two-fold reason: 
(i)   conventional RNNs are particularly difficult to train with backpropagation  because of the vanishing/exploding gradient problem; and
(ii)  RNNs' performances are often mainly due to the output weights \citep{Schiller2005}.
Thus, the main idea of reservoir computing is to use a fixed, random, large recurrent neural network, called the "reservoir", which is driven by the inputs, and  the outputs are obtained by a linear combination of the reservoir states.  A typical representation of an ESN is shown in Fig. \ref{fig:ESN}. Before going into the details, we offer a motivation from dynamical systems in Sec.~\ref{sec:ESN_motivation}.\\

\subsection{The dynamical systems' interpretation of ESNs}\label{sec:ESN_motivation}
Choosing a (good) model is an exciting activity. It takes courage to make assumptions, domain knowledge to justify the assumptions, rigour to translate the assumptions into mathematics, and creativity to combine it all. A good model is a model that is able to accurately predict a quantity in an unseen scenario.   
We offer a principled interpretation of ESNs as appropriate function approximators of the dynamical equations~\eqref{eq:ML_24}. \\

We are given observations of a dynamical system in form of data, $\mathbf{x}(t_i)$ at discrete time instants $t_i$, with $i=0,1,\ldots,N-1$. Observations are the {\it effects} (observables) of some unknown {\it causes} (dynamical equations) that act on some unknown {\it states}. 
 This is the starting point\footnote{This starting point is the opposite to an equation-based starting point of solving~\eqref{eq:system} from initial conditions, in which we know the {\it causes} and we need to compute the {\it effects.}}. 
First, 
we assume that the observations (data) that we see, $\mathbf{x}(t_i)$ are only a projection of a higher dimensional dynamical system (the state, $\mathbf{r}(t_i)$)
\begin{align}\label{eq:4.27}
\mathbf{x}(t_i) = \mathbf{A}\mathbf{r}(t_i), 
\end{align}
where $\mathbf{A}$ is a wide rectangular matrix, which is a projector. We will call the high-dimensional state, $\mathbf{r}(t_i)$, the {\it reservoir state}. This is a key step in reservoir computing. Why do we wish to add dimensions to the system, which seems to be an unattractive feature at a first glance? The answer is simple: More dimensions means more freedom (to make errors). More freedom means more exploration. More exploration means more learning. More learning means more accuracy. 
%
Second, we need to prescribe how the unknown reservoir state evolves in time. We do not know the equations, therefore, we need to come up with some dynamical law ourselves. Where to start? We draw on the formal solution of dynamical systems~\eqref{eq:ML_26}, in which the future state is a nonlinear function of the present and, recursively, of the past. Thus, we prescribe 
\begin{align}
\mathbf{r}(t_{i+1}) = 
\mathbf{r}(t_{i}) + \mathbf{G}\left(\mathbf{r}(t_{i})\right). 
\end{align}
Because we are taking a data-driven modelling approach and we might not know the equations, we probably know nothing about the nonlinear transformation $\mathbf{G}$. This is indeed a users' choice (ans\"atz), which brings us to the next point. 
Third, we choose our ans\"atz. 
We assume that we have set up a reservoir state that is much larger than the actual physical state (that we do not know). Therefore, any component of the reservoir state will  affect only a handful of other components\footnote{
If a component were to affect all the other components, it would mean that all the components would affect the state and, therefore, the observables. This is in contradiction with the assumption that our reservoir state is much larger than the actual state.}.
Mathematically, this modelling decision translates to 
\begin{align} \label{eq:r243un4738943}
\mathbf{r}(t_{i+1}) = 
\mathbf{r}(t_{i}) + g\left(\mathbf{W}\mathbf{r}(t_{i})\right), 
\end{align}
where $\mathbf{W}$ is a sparsely connected square matrix (more details in Sec.~\ref{sec:ESN_architecture}), and $g$ is an element-wise nonlinearity.  
Fourth, we need to connect the reservoir state's dynamical equation~\eqref{eq:r243un4738943} with the observables. 
We might be tempted to use the projection equation~\eqref{eq:4.27}, but we cannot because $\mathbf{A}$ is unknown. However, equation~\eqref{eq:4.27} tells us that there exists an infinite number of reservoir states that have the same observables (this is because the matrix is wide rectangular, and assumed fully ranked). This is the beauty of working in higher-dimensional spaces: We have a good amount of (in fact, infinite) freedom to describe the observables (or equivalently, we have a good amount of freedom to make mistakes and rectify them at the end). Therefore, we just need to embed the observables in the reservoir space
\begin{align}
    \mathbf{r}(t_{i+1}) = 
\mathbf{W}_{in}\mathbf{x}(t_i) + g\left(\mathbf{W}\mathbf{r}(t_{i})\right), 
\end{align}
where the matrix $\mathbf{W}_{in}$ is tall. 
The  purpose of $\mathbf{W}_{in}$ is to represent the observable in a higher-dimensional space for consistency with the dimensions. 
Fifth, and finally, in traditional ESNs,  we put all the arguments inside the nonlinearity
\begin{align}\label{eq:4.31}
    \mathbf{r}(t_{i+1}) = g\left(
\mathbf{W}_{in}\mathbf{x}(t_i) + \mathbf{W}\mathbf{r}(t_{i})\right).
\end{align}
This step is not strictly necessary, but it is customary because nonlinearly transforming the data can give more expressivity to the network. Eqaution~\eqref{eq:4.31} can also be interpreted in an alternative dynamical way
\begin{align}
    \mathbf{r}(t_{i+1}) = g( \underbrace{\mathrm{forcing}}_{\sim\mathbf{x}(t_i)}; \;\;\;\underbrace{\mathrm{state}}_{\sim \mathbf{r}(t_{i})}).
\end{align}
The data plays the role of an instantaneous forcing term in the dynamical equation (to nudge the state to the observations),  the state plays the role of carrying memory of the past, and the nonlinear law $g$ is the dynamical law. 
Eq.~\eqref{eq:4.31} tells us that the future depends on our present observation and on the reservoir state, which carries memory of the past observations. We have a well-motivated ans\"atz to work with, and now we can go into the technical details.

\subsection{Architecture}\label{sec:ESN_architecture}
In its simplest form, an ESN is composed of three parts: the input layer, the dynamical reservoir and the output layer~\citep{Lukosevicius2009,lukovsevivcius2012practical}. In contrast to conventional recurrent neural networks, the weights of the input layer and the adjacency matrix in the reservoir are fixed and only the output layer is trained.  

\begin{figure}
    \centering
    \includegraphics[width=\textwidth]{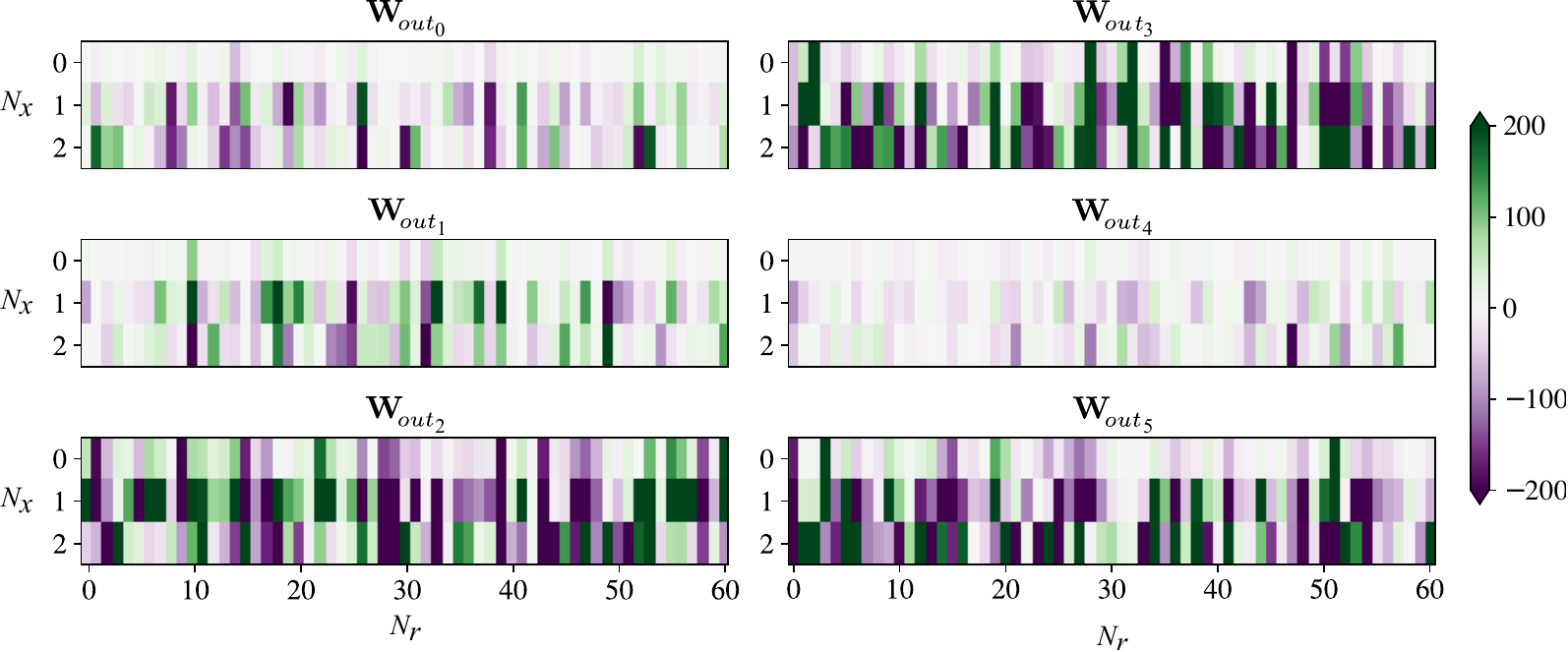}
    \caption{Different ESN initializations have different optimal hyperparameters ($\sigma_\mathrm{in}, \rho, \gamma$) and parameters (obtained after training), which are the components of the readout matrix, $\mathbf{W}_{out}$. The regularization parameter ranges between $10^{-9}$ and $10^{-12}$. }
    \label{fig:enter-fr32f3r}
\end{figure}

\begin{enumerate}
\item \textbf{Input layer}. The input layer takes the input, $\mathbf{x}(t_i)\in\mathbb{R}^{N_x}$ which is physical state, into a higher dimensional space. 
 The readout layer is represented by a tall rectangular matrix $\mathbf{W}_{in}\in \mathbb{R}^{N_r \times N_x}$, where $N_r\gg N_x$ is the reservoir's dimension.   The input matrix,
$\mathbf{W}_{in}$, has only one element different from zero per row, which is sampled from a uniform distribution in $[-\sigma_\mathrm{in},\sigma_\mathrm{in}]$, where $\sigma_\mathrm{in}$ is the input scaling. The value of $\sigma_\mathrm{in}$   indicates the sensitivity of the reservoir neurons to the input excitation and tunes the amount of nonlinearity (through the saturation of the activation function) in the reservoir. This is typically a sparse matrix in which each neuron is connected to a small number of inputs or even only to one. For tasks where there are extremely different sensitivities to the inputs, each column of $\mathbf{W}_{in}$ can be scaled differently resulting in $N_x$ different input scalings. 

\item \textbf{Reservoir}. The reservoir is the higher dimensional space in which we learn the chaotic dynamics of the physical system. Think of a reservoir as a large ``repository'' of dynamics: When you do not know the dynamics in the physical space, it is easier to work in a higher dimensional space. The reservoir retains memory of the past, which is key to learning dynamics from data. 
The reservoir is composed of $N_r$ neurons,  which are connected through an adjacency (also known as recurrent) matrix $\mathbf{W}\in \mathbb{R}^{N_r \times N_r}$. In general, the reservoir is (i) large, (ii) sparse,  (iii) randomly connected, and (iv) fixed a priori.   The components of the reservoir state, $\mathbf{r}$, also known as echoes or neurons, evolve according to
\begin{equation}
\label{eq:dyn_ESN}
\mathbf{r}(t_{i+1}) = \tanh \left(\sigma_\mathrm{in} \mathbf{W}_{in} \mathbf{x}(t_i) + \rho\mathbf{W} \mathbf{r}(t_i) \right), 
\end{equation}
in which  $\tanh$ is chosen as the nonlinearity $g$. 
Equation \eqref{eq:dyn_ESN} is the central equation governing the dynamics of the reservoir.  Typically, the adjacency matrix $\mathbf{W}$ is initialized as a sparse matrix with an average connectivity $\langle d \rangle$ with the non-zero elements being sampled from a uniform distribution between $[-1,1]$. 

The matrix $\mathbf{W}$ is rescaled to ensure that its spectral radius is equal to a prescribed value $\rho < 1$, which is a hyperparameter. The value of $\rho$ governs the amount of memory and nonlinearity in the reservoir. Observations have shown that for systems with long memory, it is preferable to use values of $\rho$ close to unity.   Having a sparse matrix improves the scalability of the ESN in terms of computational time, as the cost of the network estimation for sparse network scales with $N_r$, and not with $N_r^2$ as it would be in dense networks  \cite{Lukosevicius2009}.

\begin{tcolorbox}[breakable, opacityframe=.1, title=Echo state property]
An important property that $\mathbf{W}$ should have is the {\it echo state property}  \citep{Lukosevicius2009}. We wish that the effect of a state $\mathbf{r}(t_i)$ and an  input $\mathbf{x}(t_i)$ on a future state $\mathbf{x}(t_{i+k})$ vanishes gradually as time passes (i.e., for $k\gg 1$). This is because there are temporal correlations in dynamical systems, which, albeit they are system dependent,  have typically a finite time scale. In other words, not all past influences the present: we want to forget what does not matter to predict the future. A sufficient (but not necessary) condition to achieve this, in reservoirs with $\tanh$ activation and zero input, is to ensure that the spectral radius, which is the largest absolute value of the eigenvalues of $\mathbf{W}$, is smaller than unity\footnote{In some tasks, some reservoirs with spectral radius larger than unity may also perform well \citep{Lukosevicius2009}.}.  
\end{tcolorbox}

\item \textbf{Readout layer}. The task of the readout layer is to bring the higher-dimensional reservoir state down to the lower-dimensional physical space, where the data lives.  It is represented by a matrix $\mathbf{W}_{out}\in\mathbb{R}^{N_x \times N_r}$, which is the only trainable set of parameters of the entire network. This provides the prediction from the reservoir state as a simple linear combination 
\begin{equation}
\hat{\mathbf{x}}(t_i) = \mathbf{W}_{out} \mathbf{r}(t_i)
\end{equation}
The readout matrix, $\mathbf{W}_{out}$, is obtained from training the echo state network, as explained in Sec~\ref{sec:ESN_training}. This matrix plays the role of the projector, as per our modelling decision in~\eqref{eq:4.27} (see also Figure~\ref{fig:enter-fr32f3r}).  
\end{enumerate}

\begin{figure}[!htb]
	\centering 
 \includegraphics[width=\textwidth]{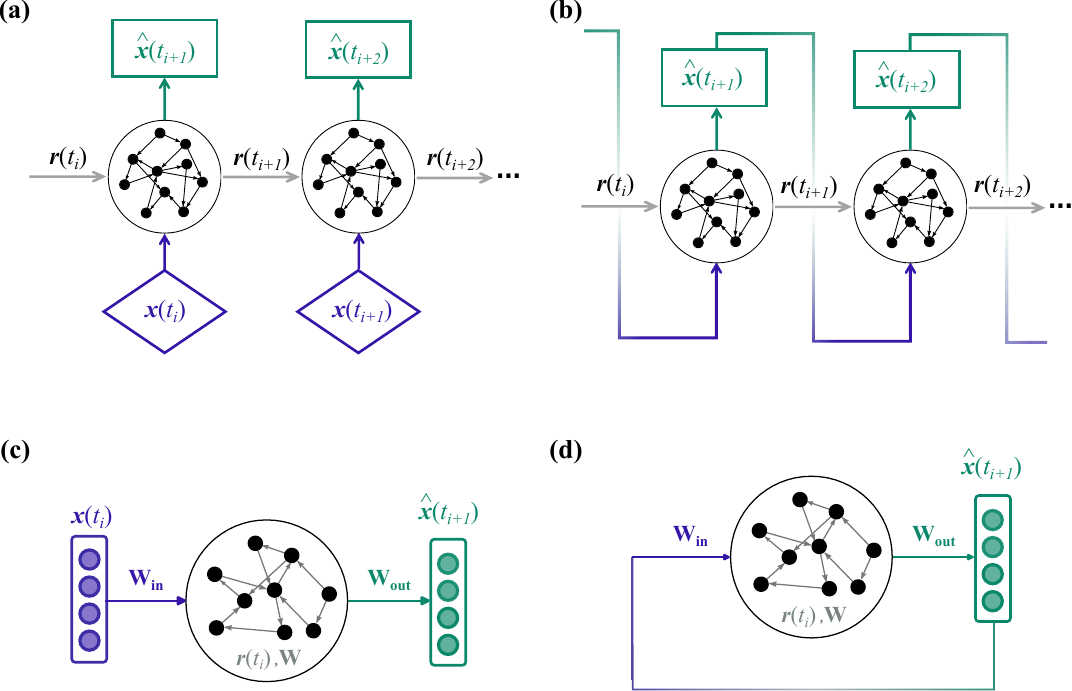}\label{fig:ESN_oloop_cloop}
	\caption{Typical Echo State Network architectures. Open-loop configuration: Unfolded representation (a), compact representation (c). Closed-loop configuration: Unfolded representation (b), compact representation (d).}
	\label{fig:ESN}
\end{figure}

\subsection{Training}\label{sec:ESN_training}
\begin{figure}[!htb]
    \centering
    \includegraphics[width=\textwidth]{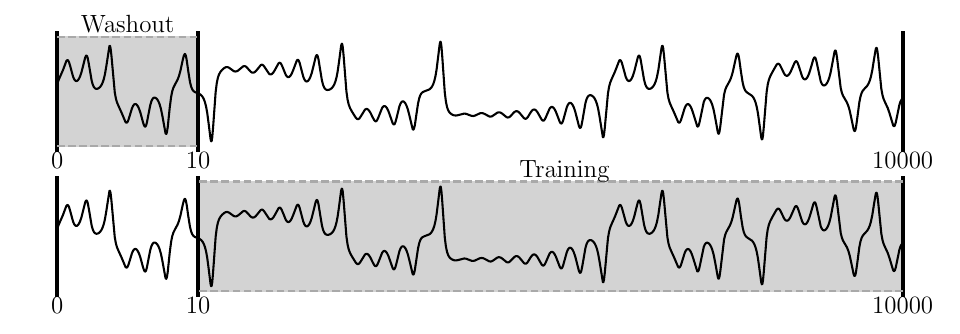}
    \caption{Split of the input data for the Echo State Network (ESN). During the washout phase (top), the ESN output is ignored and only during the training phase (bottom), the ESN is trained.}
    \label{fig:esn_washout}
\end{figure} 

The ESN can be run either in open-loop or closed-loop (Figure~\ref{fig:ESN_oloop_cloop}). 
In the open-loop configuration, which we use for training (Fig. \ref{fig:ESN_oloop_cloop}a,c), we feed the data as the input at each time step to compute and store the reservoir dynamics, $\mathbf{r}(t_i)$. 
In the initial transient of this process, which is the washout interval, we do not compute the output, $\hat{\textbf{x}}(t_i)$ (Figure~\ref{fig:esn_washout}). 
The purpose of the washout interval is for the reservoir state to satisfy the echo state property. In doing so the reservoir state becomes (i) up-to-date with respect to the current state of the system, and (ii) independent of the arbitrarily chosen initial condition, $\textbf{x}(t_0) = {0}$. 
After washout, we train the output matrix, $\mathbf{W}_{out}$. During training, we add Gaussian noise, $\mathcal{N}$, to the training inputs, $\mathbf{x}$, so that the $j$th component of the input becomes $x_j(t_i) = x_j(t_i) + \mathcal{N}(0,k_n\sigma(x_j))$, where $\sigma(\cdot)$ is the standard deviation and the input noise, $k_n$, is a tunable parameter. Adding noise to the training data improves the forecasting of chaotic dynamics with ESNs because the networks explore the region around the attractor, thereby becoming more robust to closed-loop prediction errors \citep{lukovsevivcius2012practical, vlachas2020backpropagation,racca2021robust,racca2022data}. \\ 

The readout matrix, $\mathbf{W}_{out}$, is obtained by minimising the mean squared error (MSE) between the predictions, $\hat{\textbf{x}}$, and the data, $\textbf{x}$, over the training set of $N$ points
\begin{equation}
\label{eq:MSE_eq}
    \textrm{MSE}(\hat{\mathbf{x}},{\mathbf{x}}) \equiv \frac{1}{N}
    \sum_{i=0}^{N} || \hat{\mathbf{x}}(t_i) - {\mathbf{x}}(t_i)||^2 + \frac{\gamma}{N_x}\sum_{j=0}^{N_x} ||\mathbf{w}_{out,j} ||^2, 
 \end{equation}
 where the first part is the MSE between the data available and the ESN prediction, $\mathbf{w}_{out,{j}}$ is the $j$-th row of $\mathbf{W}_{out}$ and $||\cdot  ||$ is the $l_2$-norm. The loss function~\eqref{eq:MSE_eq} represents a quadratic optimisation problem. This is excellent news: The minimum is unique and global. 
 The Tikhonov regularisation factor~\citep{tikhonov2013numerical}, $\gamma ||\mathbf{w}_{out,{j}}||^2$, penalizes large values in $\mathbf{W}_{out}$, which prevents the risk of overfitting and allows for better numerical stability in generative (i.e., closed-loop) mode\footnote{We do not know want too large values in the readout matrix, $\mathbf{W}_{out}$, because the prediction would be very sensitive to small differences in the reservoir states. This could lead to an unstable reservoir when running the reservoir in a closed-loop.}. Furthermore, the $\gamma$ factor acts as a balancing factor between fitting the data and avoiding large values in $\mathbf{W}_{out}$. In general, this term should   small  whilst ensuring that the reservoir remains stable with mitigated  overfitting.
Thanks to the output being a linear function of the reservoir state, training the network does boils down to  solving a linear system (ridge  regression) 
\begin{equation}
\label{eq:RidgeReg}
    (\mathbf{R}\mathbf{R}^\mathrm{T} + \gamma \mathbf{I})\mathbf{W}^\mathrm{T}_{out} = \mathbf{R} {\mathbf{X}}^\mathrm{T},
\end{equation}
where $\mathbf{R}\in\mathbb{R}^{N_r\times N}$ and ${\mathbf{X}}\in\mathbb{R}^{N_x\times N}$ are the horizontal concatenation of the reservoir states, and of the training data, respectively; and 
$\mathbf{I}\in\mathbb{R}^{N_r\times N_r}$ is the identity matrix. The derivation of \eqref{eq:RidgeReg} is included in Sec.~\ref{app:RidgeReg}.
The linear system~\eqref{eq:RidgeReg} can be solved with the {\tt linalg.solve} function in NumPy \citep{harris2020array}, which is a robust and numerical solution method. An alternative is given by the {\tt ridge regression} function in Scipy \citep{2020SciPy-NMeth}.

\begin{tcolorbox}[breakable, opacityframe=.1, title=Pseudoinverse]
The optimal $\mathbf{W}_{out}$, which minimizes the MSE, can be obtained analytically with a pseudoinverse 
\begin{align}
    \mathbf{W}^\mathrm{T}_{out} = (\mathbf{R}\mathbf{R}^\mathrm{T} + \gamma \mathbf{I})^{-1}\mathbf{R} \mathbf{X}^\mathrm{T}. 
    \end{align}
 Algebraically, it can be seen that the regularisation factor $\gamma$ improves the conditioning before the matrix-inversion operation. However, the pseudoinverse approach generally works for small reservoirs, and  is computationally and memory-wise expensive to perform.
\end{tcolorbox}





\begin{tcolorbox}[breakable, opacityframe=.1, title=Algorithm: ESN Training] 
\textbf{Input:} Observations in training data set: $\mathbf{X} = \begin{bmatrix} \mathbf{x}_1, ..., \mathbf{x}_{N}
\end{bmatrix}$ \\
\textbf{Parameters:} Number of washout steps $N_{washout}$
    \begin{enumerate}
        \item \textbf{Compute the reservoir for the training data.} Evolve the reservoir from Eq.~\eqref{eq:dyn_ESN} for $t_i$, with $i=1, \dots, N$ , $\mathbf{r}(t_i+1) = g \left( \mathbf{W}_{in} \mathbf{x}(t_i) + \mathbf{W} \mathbf{r}(t_i) \right)$. 
        \item \textbf{Discard $N_{washout}$ steps from reservoir and observations.}
        \item \textbf{Collect the remaining reservoir observation in the matrix $\mathbf{R}$. }
        \item \textbf{Compute $\mathbf{W}_{out}$ using ridge regression.}
            Compute the reservoir output matrix $\mathbf{W}_{out} = \textrm{ridge regression}(\mathbf{R}, \mathbf{X}, \gamma)$.
    \end{enumerate}

    \textbf{Practical tip:} Use \href{https://scikit-learn.org/stable/modules/generated/sklearn.linear_model.Ridge.html}{scikit-learn's Ridge Regression} or for \textbf{Step 4}. Validate hyperparameters ($\rho, \sigma_{in}, \gamma$) and different connectivities and reservoir sizes conveniently with the methods provided in the GitHub repository \href{https://github.com/MagriLab/EchoStateNetwork}{EchoStateNetworks}.
    \end{tcolorbox}

\subsection{ESN variants}
There is a variety of variants for the ESN. We touch upon the most common architectures. 
\subsubsection{Biases}
The ESN equations~\eqref{eq:dyn_ESN} are symmetric.
  The dynamical evolution of the reservoir state can be written as
  \begin{align}
    \mathbf{r}(t_{i+1}) &= \tanh\left( 
    \tilde{
    \mathbf{W}
    }
    \mathbf{r}(t_i) 
    \right), \\
     \tilde{
    \mathbf{W}
    } & \equiv \sigma_\mathrm{in}\mathbf{W}_{in}\mathbf{W}_{out} + \rho\mathbf{W}.
  \end{align}
    This means that taking some reservoir state $\mathbf{r}(t_i)$, and flipping its sign, i.e., $-\mathbf{r}(t_i)$, we obtain $-\mathbf{r}(t_{i+1})$. Thus, either the ESN admits two attractors symmetric to each other, or it admits one symmetric attractor. This is something we want to avoid, which is why we break the symmetry with the bias~\citep{Huhn2020}. 
  To break the symmetry, it is customary to add biases in the inputs and outputs layers~\citep{lu2017reservoir,huhn2020learning}.  The input bias is a hyperparameter, which is selected in order to have the same order of magnitude of the normalized inputs, while the output bias is determined by training the weights of the output matrix.
With biases, the reservoir dynamics are governed by 
\begin{equation}
\mathbf{r}(t_{i+1}) = \tanh\left( \sigma_\mathrm{in}\mathbf{W}_{in}[b_\mathrm{in}; \mathbf{x}(t_{i})] + \rho\mathbf{W}\mathbf{r}(t_{i}) \right)
\end{equation}
where $b_\mathrm{in}$ is the input bias, which is typically set to 1; and $[~;]$ indicates a vertical concatenation. The output is obtained as 
\begin{equation}
\hat{\mathbf{x}}(t_{i}) = \mathbf{W}_{out}[b_\mathrm{in}; \mathbf{r}(t_{i})],
\end{equation}
where the matrices' dimension are increased by one to accommodate the newly introduced bias' dimension. 
%


\subsubsection{Leakage}
Another commonly used variant of the standard ESN is the leaky ESN where a ``leak'' integration of the previous reservoir states is performed. In this architecture, the reservoir dynamics are governed by 
\begin{equation}
\mathbf{r}(t_{i+1}) = 
 (1-\alpha) \mathbf{r}(t_{i})
+ 
\alpha  \tanh\left( \sigma_\mathrm{in}\mathbf{W}_{in} \mathbf{x}(t_{i}) + \rho\mathbf{W} \mathbf{r}(t_{i})\right) 
\end{equation}
where $\alpha\in(0,1]$ is the leakage parameter. This approach allows to control the speed (or inertia) of the reservoir dynamics.  Small values of $\alpha$ induce a reservoir that reacts slowly (large inertia) to the input because the updated values are close to the previous. In the limiting (and pointless) case of $\alpha =0$, the reservoir does not evolve at all.  \\

\subsubsection{Physics-informed ESN (PI-ESN)}\label{app:PI_ESN_app}
 
We can embed/constrain some prior knowledge in the system in the training of the ESN~\citep{doan2020physics,doan2021short}. Assuming that the system under study is  governed by~\eqref{eq:system}, we can consider collocation points for the ESN at times $t>(N-1)\Delta t$ after the training. Then, if we consider $N_p$ collocation points, the prediction from the ESN over a time period $t \in [N\Delta t, (N+N_p-1) \Delta t]$, noted $\{ \widehat{\mathbf{x}}(n_p) \}_{n_p=1,...,N_p}$, can be collected and the physical residual is estimated as 
\begin{equation}
L_p = \frac{1}{N_p} \sum_{n_p=1}^{N_p} || \mathbf{F}(\widehat{\mathbf{x}}(n_p)) ||^2
\end{equation}
By combining the MSE and this physical residual, a new loss function can be used for the training of the ESN, which regularizes $\mathbf{W}_{out}$ with the physical residual at the  additional collocation points 
\begin{equation}
L_{phys} = \frac{1}{N} \sum_{i=0}^{N-1} ||\hat{\mathbf{x}(t_i)} - {\mathbf{x}(t_i)}  ||^2  + \gamma \frac{1}{N_p} \sum_{n_p=1}^{N_p} || \mathbf{F}(\widehat{\mathbf{x}}(n_p)) ||^2
\label{eq:lossPhys_ESN}
\end{equation}

The regularisation term in Eq.~\eqref{eq:lossPhys_ESN} acts as a physics-based regularisation factor, as compared to the Tikhonov regularisation, which acts on the norm of $\mathbf{W}_{out}$. This physics-constrained loss function improves the resulting trained ESN, the training of which can be obtained by gradient-based optimisation starting for the data-only optimal $\mathbf{W}_{out}$.

%

\subsection{Closed-loop}
\label{sec:validation}
The objective is to learn a data-driven model of a physical dynamical system. Thus, to compare the capability of the ESN to model an existing system, it is the generative performance of the ESN that has to be employed. This is known as {\it closed-loop} mode (or configuration). 
The closed-loop configuration is used for validation and testing (Fig.~\ref{fig:ESN_oloop_cloop}b,d), starting from an initial data point as an input and an initial reservoir state obtained after the washout interval, the output, $\hat{\textbf{x}}$, is fed back to the network as an input for the next time step prediction. 
In doing so, the network is able to autonomously evolve in the future. The reader is referred to~\citet{racca2021robust,racca2023neural} for an in-depth discussion of these aspects. 

\subsection{Validation}
During validation, we use part of the data to select the hyperparameters of the network by minimising an objective function, which is usually the error between the prediction and the data. 
 ESN hyperparameters belong in two categories: 
 (i) those that require re-initialisation, i.e.,  $\mathbf{W}_{\mathrm{in}}$ and $\mathbf{W}$; 
and (ii) those that do not require re-initialisation. 
The size of the reservoir, $N_r$, and connectivity, $d$, require re-initialisation, whereas the input scaling, $\sigma_{\mathrm{in}}$, the spectral radius, $\rho$, the Tikhonov parameter, $\gamma$, the input noise, $k_n$, and the input bias $b_{\mathrm{in}}$, do not. 
The fundamental difference between (i) and (ii) is that the random component of the re-initialisation of $\mathbf{W}_{\mathrm{in}}$ and $\mathbf{W}$ makes the objective function to be optimized random, which significantly increases the complexity of the optimisation. 
We therefore optimize the input scaling, $\sigma_{\mathrm{in}}$, spectral radius, $\rho$, and Tikhonov parameter, $\gamma$, which are  key hyperparameters for the performance of the network \citep{lukovsevivcius2012practical,jiang2019model}. Specifically, we explore $(\sigma_{\mathrm{in}},\rho)$ space, and perform a grid search within each evaluated $[\sigma_{\mathrm{in}},\rho]$ to select $\gamma$. This is because of the different computational cost of evaluating multiple Tikhonov parameters with respect to other hyperparameters.

\begin{figure}[ht!]
    \centering
    \includegraphics[width=1.\textwidth]{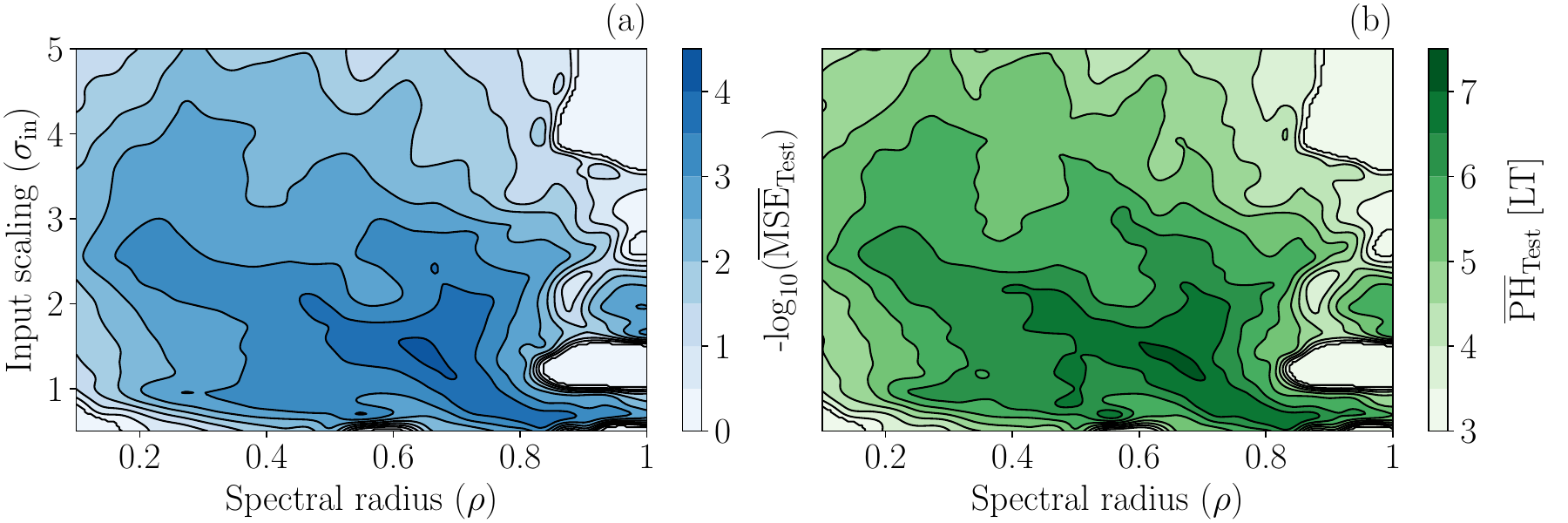}
    \caption{Mean of the Gaussian process reconstruction from a $30\times30$ grid for the average (a) MSE in 3 LTs intervals and (b) Prediction Horizon (PH) in the test set for for an echo state network in the Lorenz system. For visualisation purposes, we saturate the $\mathrm{MSE}$ to be $\leq 1$, and the PH to be $\geq 3$~\citep{racca2021robust}.
    }
    \label{fig:PH_MSE_surf}
\end{figure}

\subsubsection{Validation metrics}
We determine the hyperparameters by minimising the mean squared error \eqref{eq:MSE_eq} in validation intervals of fixed length. 
The networks are tested on multiple starting points along the attractor by using both the MSE and the prediction horizon (PH). The prediction horizon is the time interval during which the instantaneous normalized root mean squared error (NRMSE) is smaller than the user-defined threshold, $k_{PH}$ 
\begin{gather}
\label{eq:PH}
\mathrm{PH} = \mathop{\mathrm{argmax}}_{t}(t \,|\, \mathrm{NRMSE}(\mathbf{x}(t),\hat{\mathbf{x}}(t)) < k_{PH}), \\
\mathrm{NRMSE}  = \frac{1}{\mathrm{NORM}} 
\sqrt{\sum_{j=1}^{N_x} \frac{1}{N_x} (\hat{x}_j(t)-x_{j}(t))^2 }\,,
\label{eq:NRMSE}
\end{gather}
where $t$ is the time from the start of the closed-loop and $\mathrm{NORM}$ is the normalisation factor (e.g, the mean of the norm or the standard deviation of the data \citep{doan2020physics,vlachas2020backpropagation,doan2021short,racca2021robust,racca2022data}).
The prediction horizon is a commonly used metric, which is tailored for the prediction of diverging trajectories in chaotic dynamics \citep[e.g,][]{boffetta2002predictability,pathak2018hybrid}.
The mean squared error and prediction horizon for the same starting points in the attractor are correlated (Fig. \ref{fig:PH_MSE_surf}). This means that selecting the hyperparameters by minimising the MSE is analogous to maximising the prediction horizon, as shown in~\citet{racca2021robust}.
 \subsubsection{Strategies}
\label{sec:valstrat}
\begin{figure}
    \centering
    \includegraphics[width=1.0\textwidth]{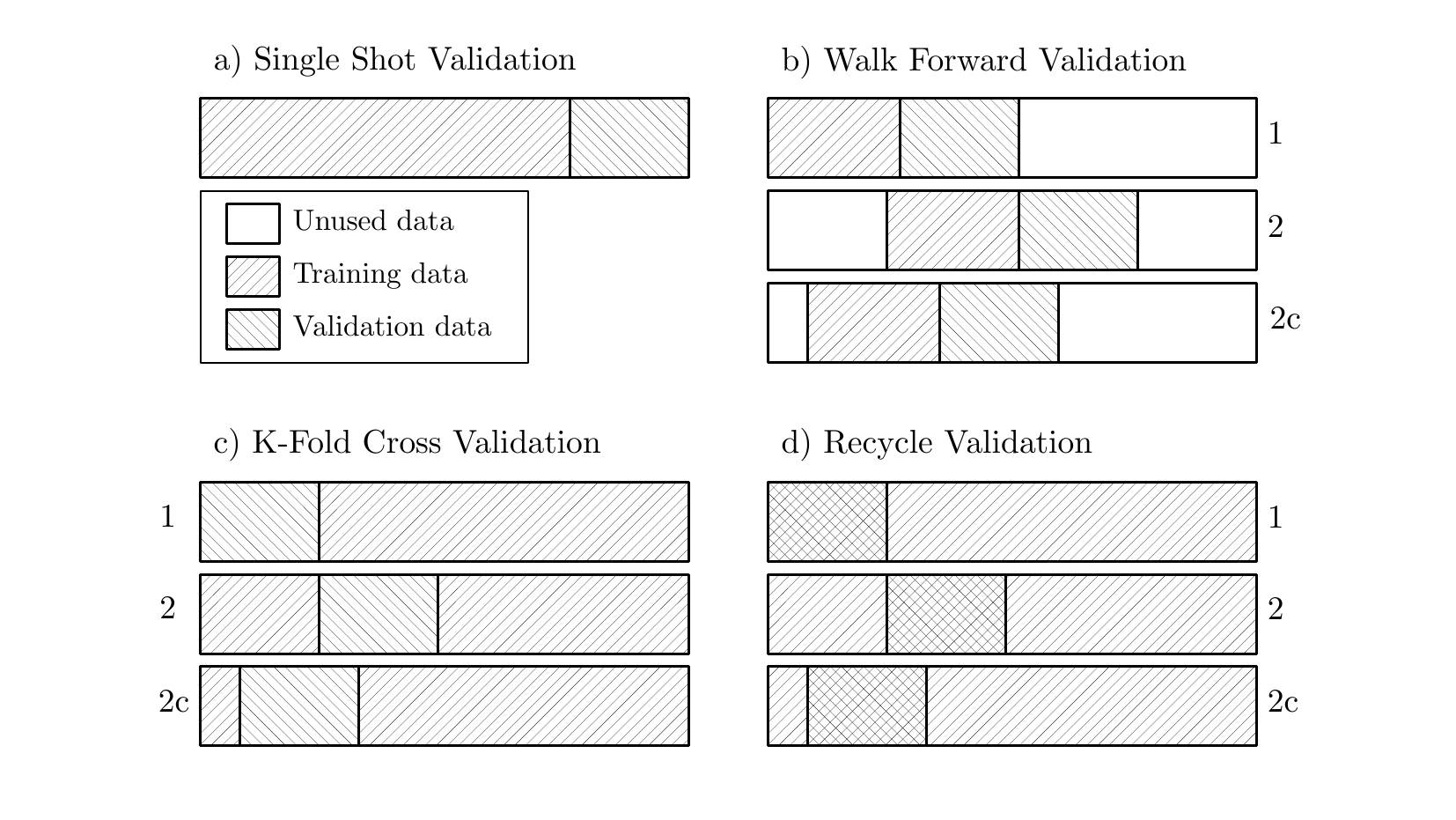}
    \caption{Partition of the data in the different validation strategies~\citep{racca2021robust}. In (b-d),  bar 1 shows the first fold,   bar 2 shows the second fold, and bar 2c shows the second fold in the chaotic version (shifted by one Lyapunov time).}
    \label{Val_Strat}
\end{figure}

Validations strategies for chaotic ESN were developed and analysed in~\citet{racca2021robust}. The validation strategy is the procedure that determines which part of the data we use for training and validation.
The most common validation strategy for ESNs is the single shot validation (SSV), which splits the available data in a training set and a single subsequent validation set (Fig.~\ref{Val_Strat}a). 
The time interval of the validation set, during which the hyperparameters are tuned, is small and represents only a fraction of the attractor. 
In time series prediction, the choice of validation strategy has to take into account (i) the intervals we are interested in predicting, and (ii) the nature of the signal we are trying to learn. 
Here, we are interested in predicting multiple intervals as the trajectory spans the attractor, rather than a specific interval starting from a specific initial condition. 
Moreover, the trajectory that spans the attractor is ergodic, e.g., there is no time-dependency of the mean of the signal, so that trajectories return indefinitely in nearby regions of the attractor (see section \ref{sec:chaos_stats}).
Thus, we can obtain information regarding the intervals that we are interested in predicting from any interval of the trajectory that constitutes our dataset, regardless of the interval position in time within the dataset. This means that (i) all the parts of the dataset are equally important in determining the hyperparameters, and (ii) the validation should be performed on the entire dataset and not only on the last portion of it.
For this reason, the single shot validation may not be well-suited to chaotic time series prediction. Robust validation strategies are described next. \\ 

\textbf{Walk forward validation.} In the walk forward validation (WFV) (Fig. \ref{Val_Strat}b), we partition the available data in multiple splits, while maintaining the sequentiality of the data. 
From a starting dataset of length $n$, the first $m$ points ($m<n$) are taken as the first fold, with $N_t$ points for training and $v$ points for validation ($v+N_t=m$). 
These quantities must respect $(n-m) = (k_1-1) v; \; k_1 \in \mathbb{N}$ to have an integer number of folds of same size.
The remaining $(k_1-1)$ folds are generated by moving the training plus validation set forward in time by a number of points $v$. 
This way, the original dataset is partitioned in $k_1$ folds and the hyperparameters are selected to minimize the average MSE over the folds. For every set of hyperparameters and every fold, the output matrix, $\mathbf{W}_{\textrm{out}}$, is recomputed. \\ 

\textbf{K-fold cross validation.} Although the K-fold cross validation (KFV) (Fig. \ref{Val_Strat}c) is a common strategy in regression and classification, it is not commonly used in time series prediction because the validation and training intervals are not sequential to each other. 
This strategy partitions the available data in $k_2$ splits.  
Over the entire dataset of length $n$, after an initial $bv$ points, with $0\leq b<1$, needed to have an integer number of splits, the remaining $n-bv$ points are used as $k_2$ validation intervals, each of length $v$. 
For each validation interval we define a different fold, in which we use all the remaining data points for training. 
We determine the hyperparameters by minimising the average of the MSE between the folds. For every set of hyperparameters and every fold, the output matrix, $\mathbf{W}_{\textrm{out}}$, is recomputed.\\ 

\textbf{Recycle validation.} \citet{racca2021robust} proposed the recycle validation (RV) (Fig. \ref{Val_Strat}d), which exploits the information obtained by both open-loop and closed-loop configurations. Because the network works in two different configurations, it can obtain additional information when validating on data already used in training. 
To do so, first, we train $\mathbf{W}_{\textrm{out}}$ only once per set of hyperparameters using the entire dataset of $n$ points.
Second, we validate the network on $k_2$ splits of length $v$ from data that has already been used to train the output weights. 
Each split is imposed by moving forward in time the previous validation interval by $v$ points. 
After an initial $bv$ points, with $0\leq b<1$, needed to have an integer number of splits, the remaining $n-bv$ points are used as $k_2$ validation intervals. 
We determine the hyperparameters by computing the average of the MSE between the splits. 
This strategy has four main advantages.
First, it can be used in small datasets, where the partition of the dataset in separate training and validation sets may cause the other strategies to perform poorly. In small datasets, the validation intervals represent a larger percentage of the dataset, since each validation interval needs to be multiple Lyapunov Times to capture the divergence of chaotic trajectories. Therefore, the training set becomes substantially smaller than the dataset and the output matrix used during validation differs substantially from the output matrix of the whole dataset. This results in a poor selection of hyperparameters.
Second, for a given dataset, we maximize the number of validation splits, using the same validation intervals of the K-fold cross validation. 
Third, we tune the hyperparameters using the same output matrix, $\mathbf{W}_{out}$, that we use in the test set. 
Fourth, it has a significantly lower computational cost than the K-fold cross validation because it does not require retraining the output matrix for the different folds. \\ 

\textbf{Chaotic version.} The chaotic version of a validation strategy consists of shifting the validation intervals forward in time, not by their own length, but by one Lyapunov time when constructing the next fold. 
In doing so, different splits will overlap, but, since the closed-loop prediction related to the split that started 1 LT earlier has strayed away from the attractor on average by $e^{\Lambda_1\times 1 \,\mathrm{LT}}=e$, the two intervals contain different information. 
The purpose of this version is to further increase the number of intervals on which the network is validated. 
The regular and chaotic versions for each validation strategy are shown in Fig. \ref{Val_Strat}b-d in bars 2 and 2c, respectively.
The chaotic versions of the walk forward validation, the K-fold cross validation and the recycle validation are denoted by the subscript $\mathrm{c}$. \\

\textbf{Computing $\mathbf{W}_{out}$}. In each validation strategy, for each combination of input scaling, $\sigma_{\mathrm{in}}$, spectral radius, $\rho$, and Tikhonov parameter, $\gamma$, we compute the output matrix, $\mathbf{W}_{out}$. Moreover, for each combination of $\sigma_{\mathrm{in}}$, $\rho$ and $\gamma$ in each fold of the K-fold validation and walk forward validation a different $\mathbf{W}_{out}$ is computed. Even with same hyperparameters the folds have different $\mathbf{W}_{out}$ because the training data is different. For each validation strategy, once $\mathbf{W}_{out}$ is determined in open-loop, the error that is minimized is that obtained by running the network in closed-loop in the validation interval(s). After training and validation are completed---i.e., we have selected the hyperparameters---the $\mathbf{W}_{out}$ to be used in the test set is computed on the entire dataset used for training plus validation using the optimal hyperparameters. \\

    \subsection{Jacobian of the ESN}
We mathematically derive the Jacobian of the Echo State Network. The reservoir's evolution equation can be recast in a compact form as 
  \begin{align}
    \mathbf{r}(t_{i+1}) &= \tanh\left( 
    \tilde{
    \mathbf{W}
    }
    \mathbf{r}(t_i) 
    \right), \\
     \tilde{
    \mathbf{W}
    } & \equiv \sigma_\mathrm{in}\mathbf{W}_{in}\mathbf{W}_{out} + \rho\mathbf{W}.
  \end{align}
 The Jacobian of the ESN reservoir  in closed-loop is the total derivative of the reservoir state  at a single timestep~\citep{margazoglou2023stability} 
 \begin{align}
 \mathbf{J}(t_{i}) &\equiv \frac{\mathrm{d}\mathbf{r}(t_{i+1})}{\mathrm{d}\mathbf{r}(t_{i})} \nonumber\\
 & = \frac{\mathrm{d}\tanh\left( 
    \tilde{
    \mathbf{W}
    }
    \mathbf{r}(t_i) 
    \right)}{\mathrm{d}\mathbf{r}(t_{i})} \nonumber \\
    & = \frac{\mathrm{d}\tanh\left( 
    \tilde{
    \mathbf{W}
    }
    \mathbf{r}(t_i) 
    \right)}{\mathrm{d}\left( 
    \tilde{
    \mathbf{W}
    }
    \mathbf{r}(t_i) 
    \right)}
    \frac{
    \mathrm{d} 
    \left( 
    \tilde{
    \mathbf{W}
    }
    \mathbf{r}(t_i) 
    \right)}{\mathrm{d}\mathbf{r}(t_i)} \nonumber \\ 
    &=
   (1-\tanh^2(\tilde{
    \mathbf{W}
    }
    \mathbf{r}(t_i)))
    \tilde{\mathbf{W}}^\mathrm{T}.
 \end{align}

The Jacobian of the ESN is cheap to calculate as the expression $\left( \mathbf{W}^\mathrm{T}_{out}\mathbf{W}^\mathrm{T}_{in} + \mathbf{W}^\mathrm{T} \right)$ is a constant matrix, which is fixed after the training of $\mathbf{W}_{out}$.


\section{Long short-term memory network}\label{sec:lstms}
%
Long short-term memory networks (LSTMs) were introduced in \cite{Hochreiter_1997_LongShortTermMemory} as a type of RNN that maintains different memories for long and short-term inputs. These networks feature an architecture with a cell state, responsible for retaining long-term information, and a hidden state, focused on capturing short-term memory. The information flow within the LSTM is controlled by gating mechanisms, namely input, forget and output gates. These gates also mitigate the vanishing gradient problem caused by the backpropagation through long recurrences.

\subsection{Architecture}

The network is characterized by a cell state $\mathbf{c}_{i} \in \mathbb{R}^{N_{h}}$ and a hidden state $\mathbf{h}_{i} \in \mathbb{R}^{N_{h}}$, both of dimensions $N_h \in \mathbb{R}$, which are updated at each recurrent step. Giving an input $\mathbf{x}_{in}(t_i)$, the three gates are computed.

\begin{enumerate}
    \item \textbf{Input Gate}. The input gate determines which information from the current input $\mathbf{x}_{in}(t_i)$ should be stored in the cell state. It is defined as
    \begin{equation*}
         \mathbf{i}_{i+1} = \sigma \left(\mathbf{W}^i [\mathbf{x}_{in}(t_i); \mathbf{h}_{i}] + \mathbf{b}^i \right), 
    \end{equation*}
    where $\mathbf{b}$ is a trainable bias, and $\sigma(\cdot)$ is the sigmoid activation function. 
    
    \item \textbf{Forget Gate}. The forget gate determines which information from the previous cell state $\mathbf{c}_i$ should be discarded or kept for the current time step, and is computed as
        \begin{equation*}
        \mathbf{f}_{i+1} = \sigma \left(\mathbf{W}^f [\mathbf{x}_{in}(t_i); \mathbf{h}_{i}] + \mathbf{b}^f \right). 
    \end{equation*}
    The sigmoid, $\sigma(\cdot)$, is the activation function of choice because it captures the cases we want: It is 0 if we want to completely forget (erase) the information, it is 1 if we want to completely retain the information, and it is in the middle for all the other intermediate cases.  
    \item \textbf{Output Gate}. The output gate determines which information from the current cell state $\mathbf{c}_i$ should be passed to the next hidden state $\mathbf{h}_{i+1}$. The gate is given by
    \begin{equation*}
        \mathbf{o}_{i+1} = \sigma \left(\mathbf{W}^o [\mathbf{x}_{in}(t_i); \mathbf{h}_{i}] + \mathbf{b}^o \right).
    \end{equation*}
\end{enumerate}
The matrices $\mathbf{W}^i,$ $ \mathbf{W}^f,$ $ \mathbf{W}^o, \in \mathbb{R}^{N_{h} \times (N_x +N_h)}$ are the weight matrices of the gates, and $\mathbf{b}^i,$ $\mathbf{b}^f,$ $\mathbf{b}^o \in \mathbb{R}^{N_{h}}$ the corresponding biases. 
\begin{figure}[!ht]
    \centering
            \scalebox{1.0}{
\begin{tikzpicture}[
    font=\sf \scriptsize,
    >=LaTeX,
    cell/.style={
        rectangle, 
        rounded corners=1mm, 
        draw,
        very thick,
        },
    operator/.style={
        circle,
        draw,
        inner sep=-0.5pt,
        minimum height =.2cm,
        },
    function/.style={
        ellipse,
        draw=gray!0,
        inner sep=1pt
        },
    ct/.style={
        ellipse,
        draw=gray!0,
        line width = .75pt,
        minimum width=1cm,
        inner sep=1pt,
        },
    gt/.style={
        diamond,
        draw,
        minimum width=4mm,
        minimum height=3mm,
        inner sep=1pt
        },
    mylabel/.style={
        font=\scriptsize\sffamily
        },
    ArrowC1/.style={
        rounded corners=.05cm,
        thick,
        },
    ArrowC2/.style={
        rounded corners=.1cm,
        thick,
        },
    ]

    \node [cell,  fill=gray!05, minimum height =4cm, minimum width=6cm] at (0,0) {} ;

    \node [gt] (ibox1) at (-2,-0.75) {$\sigma$};
    \node [gt] (ibox2) at (-1.5,-0.75) {$\sigma$};
    \node [gt, minimum width=1cm] (ibox3) at (-0.5,-0.75) {Tanh};
    \node [gt] (ibox4) at (0.5,-0.75) {$\sigma$};

    \node [operator] (mux1) at (-2,1.5) {};
    \node [operator] (add1) at (-0.5,1.5) {+};
    \node [operator] (mux2) at (-0.5,0) {$\times$};
    \node [operator] (mux3) at (1.5,0) {$\times$};
    \node [function] (func1) at (1.5,0.75) {Tanh};

    \node[ct, label={[mylabel]previous cell state}] (c) at (-4.5,1.5) {$\mathbf{c}_{i}$};
    \node[ct, label={[mylabel]previous hidden state}] (h) at (-4.5,-1.5) {$\mathbf{h}_{i}$};
    \node[ct, diamond, draw=custompurple!90,  text=custompurple!90,label={[mylabel]left:input}] (x) at (-2.5,-3) {$\mathbf{x}(t_{i}) $};

    \node[ct, label={[mylabel]new cell state}] (c2) at (4.5,1.5) {$\mathbf{c}_{i+1}$};
    \node[ct, label={[mylabel]new hidden state}] (h2) at (4.5,-1.5) {$\mathbf{h}_{i+1}$};
    \node[ct, diamond, draw=customgreen!90,  text=customgreen!90, label={[mylabel]left:prediction}] (x2) at (2.5,3) {$\mathbf{\Hat{x}}(t_{i+1})$};
    \draw [ArrowC1] (c) -- (mux1) -- (add1) -- (c2);

    \draw [ArrowC2] (h) -| (ibox4);
    \draw [ArrowC1] (h -| ibox1)++(-0.5,0) -| (ibox1); 
    \draw [ArrowC1] (h -| ibox2)++(-0.5,0) -| (ibox2);
    \draw [ArrowC1] (h -| ibox3)++(-0.5,0) -| (ibox3);
    \draw [ArrowC1] (x) -- (x |- h)-| (ibox3);

    \draw [->, ArrowC2] (ibox1) -- (mux1);
    \draw [->, ArrowC2] (ibox2) |- (mux2);
    \draw [->, ArrowC2] (ibox3) -- (mux2);
    \draw [->, ArrowC2] (ibox4) |- (mux3);
    \draw [->, ArrowC2] (mux2) -- (add1);
    \draw [->, ArrowC1] (add1 -| func1)++(-0.5,0) -| (func1);
    \draw [->, ArrowC2] (func1) -- (mux3);

    \draw [-, ArrowC2] (mux3) |- (h2);
    \draw (c2 -| x2) ++(0,-0.1) coordinate (i1);
    \draw [-, ArrowC2] (h2 -| x2)++(-0.5,0) -| (i1);
    \draw [-, ArrowC2] (i1)++(0,0.2) -- (x2);

\end{tikzpicture}
}
\caption{Schematic representation of the LSTM cell structure.}
\label{fig:lstm_cell}
\end{figure}
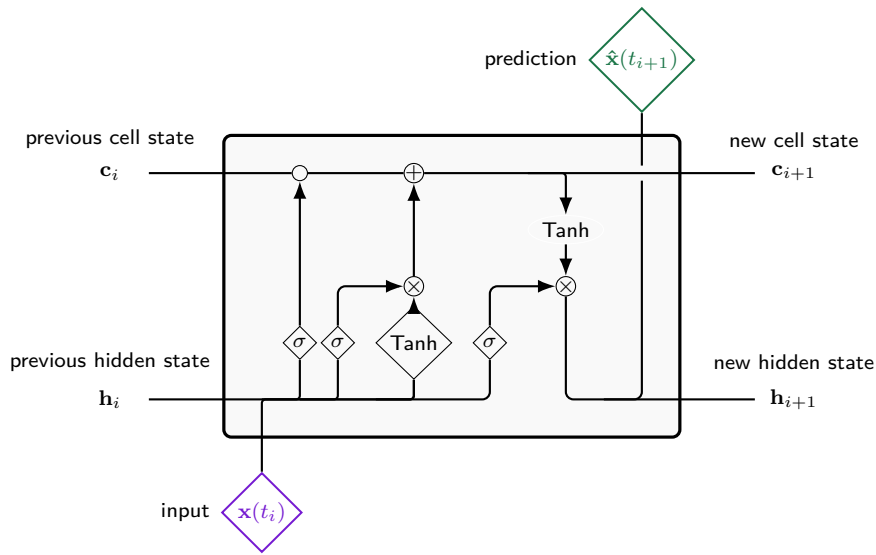
Using these gates, the next step is to compute the states of the LSTM. 
\begin{enumerate}
    \item \textbf{Cell state}. The cell state $\mathbf{c}_{i+1} $ combines the information from the input and forget gate, and corresponds to the longer memory. The state is computed in two stages
    \begin{align}
         \mathbf{\Tilde{c}}_{i+1} &= \tanh{\left(\mathbf{W}^{{g}} [\mathbf{x}_{in}(t_i); \mathbf{h}_{i}] + \mathbf{b}^{{g}} \right) }, \nonumber\\ 
    \mathbf{c}_{i+1} &= \mathbf{f}_{i+1}*\mathbf{c}_{i} + \mathbf{i}_{i+1}*\mathbf{\Tilde{c}}_{i+1}, 
    \end{align}
    where $*$ denotes the elementwise multiplication.
    \item \textbf{Hidden state}. The hidden state $ \mathbf{h}_{i+1}$ uses the information of the cell state, together with the output gate. The hidden state is directly used for the prediction and therefore corresponds to the short-term memory, i.e.
    \begin{align*}
        \mathbf{h}_{i+1} &= \tanh{\left(\mathbf{c}_{i+1} \right)} * \mathbf{o}_{i+1}.
    \end{align*}
\end{enumerate}

The hidden state is fed through a dense layer to compute the network prediction
  \begin{align*}
     \mathbf{\Hat{x}}(t_{i+1}) &= \mathbf{W}^{dense} \mathbf{h}_{i+1} + \mathbf{b}^{dense}, 
  \end{align*}
  with $\mathbf{W}^{dense} \in \mathbb{R}^{N_x\times N_h}$  and $ \mathbf{b}^{dense} \in \mathbb{R}^{N_x} $.
The weights and biases of the LSTM are trained using backpropagation through time, a process that iteratively minimizes the loss function by computing the gradient with respect to the parameters. Despite its higher time intensity compared to least-squares regression, this training method is essential for the LSTM, which typically operates with a significantly smaller hidden state dimension than ESNs (Sec.~\ref{sec:esns}) to effectively propagate the dynamics. \\
\subsection{Closed-loop}
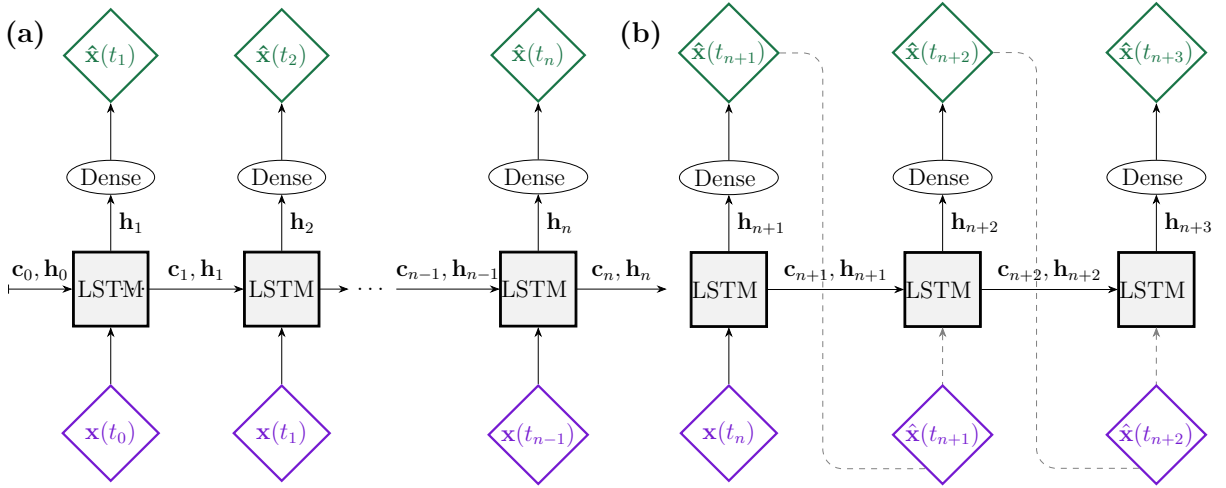
\begin{figure}[!ht]
    \scalebox{0.75}{
  \begin{tikzpicture}[x=3cm, y=1.5cm, >=Stealth]
    \foreach \jlabel [count=\j, evaluate={\k=int(mod(\j-1,1)); \jj=int(\j-1);}]
      in {0, 1}{
        \foreach \ilabel [count=\i] in {1}
            \node [neuron, fill=gray!10, align=left] at (\j, 1-\i) (h-\i-\j){ LSTM};   
          \node [fit=(h-1-\j) (h-1-\j), inner sep=0, draw] (b-\j){ } ;
          \node [input_io, below=of b-\j] (v-\j) {$ \mathbf{x}(t_{ \jlabel})$};
        \node [dense, above=of h-1-\j, align=left] (d-\j) {{{Dense}}};
          \draw [->] (v-\j) -- (b-\j);
          \draw [->] (b-\j.north) -- (d-\j) node [midway, right] {$\mathbf{h}_{\j}$};
          }
          \draw (0.5,3) node {\large{\textbf{(a)}}};
        \node [output_io, above=of d-1] (output1){$\mathbf{\Hat{x}}(t_{1})$ };
         \node [output_io, above=of d-2] (output2){$\mathbf{\Hat{x}}(t_{2})$ };
        \node [right=0.5cm of h-1-2](dots1) {\ldots};
        \node [neuron, fill=gray!10,  align=left] at (3.5, 0) (h-1-4){LSTM 
            };
        \node [fit=(h-1-4) (h-1-4), inner sep=0, draw] (b-4){} ;
        \node [input_io, below=of b-4] (v-4) {$  \mathbf{x}(t_{ n-1})  $};
         \node [right=1=0.5cm of v-2](dots2) {\ldots};
        \draw [->] (v-4) -- (b-4);
        \draw [->] (h-1-1.east) -- (h-1-2.west)node [midway, above] {$\mathbf{c}_1, \mathbf{h}_{1}$} ;
        \draw [->] (h-1-2.east) -- (dots1.west)node [midway, above] {};
        \draw [->] (dots1.east) -- (h-1-4.west)node [midway, above] {$\mathbf{c}_{n-1}, \mathbf{h}_{n-1}$};
         \draw [->] (h-1-4.east) -- (4.25, 0 )node [midway, above] {$\mathbf{c}_{n}, \mathbf{h}_{n}$};
        \node [dense, above=of b-4, align=left] (d-4) {{{Dense}} };
        \draw [->] (b-4.north) -- (d-4.south)node [midway, right] {$\mathbf{h}_{n}$} ;
        \node [output_io,  above=of d-4] (output){$\mathbf{\Hat{x}}(t_{ n})$ };
        \draw [->] (d-4.north) -- (output.south);
        \draw [->] (d-1.north) -- (output1.south);
        \draw [->] (d-2.north) -- (output2.south);
     \draw[|->] (0.4,0) -- (b-1.west)node [midway, above] {$\mathbf{c}_0, \mathbf{h}_{0}$};
    \draw (4.1,3) node {\large{\textbf{(b)}}};
     
    \end{tikzpicture}   \begin{tikzpicture}[x=3cm, y=1.5cm, >=Stealth]
       
        \node [neuron, fill=gray!10, align=left] at (3.5, 0) (h-1-4){LSTM 
            };
        \node [fit=(h-1-4) (h-1-4), inner sep=0, draw] (b-4){} ;
        \node [input_io, below=of h-1-4] (v-4) {$  \mathbf{x}(t_{ n})  $};
        \draw [->] (v-4) -- (h-1-4);
        \node [dense, above=of h-1-4, align=left] (d-4) {{{Dense}} };
        \draw [->] (h-1-4.north) -- (d-4.south)node [midway, right] {$\mathbf{h}_{n+1}$} ;
        \node [output_io, above=of d-4] (output){$\mathbf{\Hat{x}}(t_{ n+1})$ };
        \draw [->] (d-4.north) -- (output.south);
        \node [neuron, fill=gray!10, align=left] at (4.75, 0) (h-1-6){LSTM 
            };
        \node [fit=(h-1-6) (h-1-6), inner sep=0, draw] (b-6){} ;
        
         \draw[->, rounded corners=10pt, gray, dashed] (output.east) -- ++(0.25, 0) -- ++(0, -4.9) -- ++(0.7, 0) -- (h-1-6.south);
        \node [input_io, fill=white, below=of h-1-6] (v-6) {$  \Hat{\mathbf{x}}(t_{ n+1})  $};
    \draw [->] (h-1-4.east) -- (h-1-6.west)node [midway, above] {$\mathbf{c}_{n+1}, \mathbf{h}_{n+1}$};
        \node [dense, above=of b-6, align=left] (d-6) {{{Dense}} };
        \draw [->] (b-6.north) -- (d-6.south)node [midway, right] {$\mathbf{h}_{n+2}$} ;
        \node [output_io, above=of d-6] (output_n){$         \mathbf{\Hat{x}}(t_{ n+2})$ };
       \draw [->] (d-6.north) -- (output_n.south);
        \node [neuron, fill=gray!10, align=left] at (6., 0) (h-1-7){LSTM 
            };
        \node [fit=(h-1-7) (h-1-7), inner sep=0, draw] (b-7){} ;        
         \draw[->, rounded corners=10pt, gray, dashed] (output_n.east) -- ++(0.25, 0) -- ++(0, -4.9) -- ++(0.7, 0) -- (h-1-7.south);
        \node [input_io, fill=white, below=of h-1-7] (v-7) {$  \Hat{\mathbf{x}}(t_{ n+2})  $};
    \draw [->] (h-1-6.east) -- (h-1-7.west)node [midway, above] {$\mathbf{c}_{n+2}, \mathbf{h}_{n+2}$};
        \node [dense, above=of b-7, align=left] (d-7) {{{Dense}} };
        \draw [->] (b-7.north) -- (d-7.south)node [midway, right] {$\mathbf{h}_{n+3}$} ;
        \node [output_io, above=of d-7] (output_n1){$         \mathbf{\Hat{x}}(t_{n+3})$ };
       \draw [->] (d-7.north) -- (output_n1.south);
       
            
    \end{tikzpicture}}
    \caption{LSTM in open-loop configuration (a) and in closed-loop configuration (b)}
    \label{fig:lstm_network_loops}
    \end{figure}

During training and validation, the network operates in an open-loop configuration, depicted in Figure~\ref{fig:lstm_network_loops}(a). The LSTM's output at each time step receives the reference data inputs within the time window, and the LSTM states from previous cells. The states are reset to zero at the beginning of each time window. After the training, the network's weights and biases are fixed and it is operated in closed-loop mode, see Figure~\ref{fig:lstm_network_loops}(b). Following a one-time window warm-up in open-loop, the network's prediction serves as an input for the next cell. This enables the long-term prediction of the LSTM, even in the absence of data.

\subsection{Physics-informed architecture (PI-LSTM)}\label{sec:pi_lstm}

Incorporating physical knowledge of the governing equations has shown success in feed-forward neural networks, where automatic differentiation can be exploited to accurately compute the derivative of the governing equations \citep{Lagaris1998, raissi2019physics}. Compared to feed-forward neural networks, the recurrent structure of LSTM does not allow for a straightforward computation of the temporal derivative. Therefore, physics constraints in the loss function have to account for the temporal structure of the LSTM architecture. A robust approach is given by reformulating \eqref{eq:system} through the integral formulation~\eqref{eq:ML_24}, which we repeat here for clarity 
\begin{align}\label{integral_equation_dynamical_system}
\mathbf{x}(t_{i+1}) &= \int_{t_0}^{t_{i+1}} \mathbf{F}(\mathbf{x}(t)) dt \
= \mathbf{x}(t_i) + \int_{t_i}^{t_{i+1}} \mathbf{F}(\mathbf{x}(t)) dt, \ i\geq0,
\end{align}
which enables the use of numerical quadrature methods to approximate the integral $\int_{t_i}^{t_{i+1}} \mathbf{F}(\mathbf{x}(t)) dt$, instead of approximating the derivative $\frac{d\mathbf{x}(t)}{dt}$~\citep{ozalp2023physics}. The integral formulation~\citep{oezalp_chaos_2023} is compatible with explicit numerical schemes of different orders of accuracy, such as the Runge-Kutta methods.
To enforce Eq. \eqref{integral_equation_dynamical_system}, we define the residual of the dynamical system
\begin{align}
    \mathcal{R}(\mathbf{x}(t_{i+1})) = \mathbf{x}(t_{i+1}) - \left( \mathbf{x}(t_i) + \int_{t_i}^{t_{i+1}} \mathbf{F}(\mathbf{x}(t)) dt\right).
\end{align}
The solution of the dynamical system is such that $\mathcal{R}(\mathbf{x}(t_i)) = 0$ for all $t_i>t_0$. By minimizing the physics-informed loss for $N$  training points
\begin{align}
    \mathcal{L}_{pi}(\mathbf{\Hat{x}}) = \frac{1}{N} \sum_{i=0}^{N -1} \|   \mathcal{R}(\Hat{\mathbf{x}}(t_{i+1})) \|^2,
\end{align}
the network prediction is constrained to fulfil the governing equations. This loss is particularly advantageous when only partial observations of the system are available as it only constrains the network output, as opposed to data-driven losses~\citep{oezalp_chaos_2023}. Additionally, a data-driven loss can be computed between the prediction $\mathbf{\Hat{x}}(t_{i+1})$ and the training label $\mathbf{x}(t_{i+1})$
\begin{align}\label{dd_loss}
    \mathcal{L}_{dd}(\mathbf{x}, \mathbf{\Hat{x}})= \frac{1}{N} \sum_{i=1}^{N} \| \mathbf{x}(t_{i}) - \mathbf{\Hat{x}}(t_{i}) \|^2.
\end{align}
By combining the data-driven loss and weighing the physics-informed loss, the PI-LSTM is constrained using 
\begin{equation}
    \mathcal{L}(\mathbf{x}, \mathbf{\Hat{x}}) = \mathcal{L}_{dd}(\mathbf{x}, \mathbf{\Hat{x}})  + \alpha_{pi}\mathcal{L}_{pi}(\mathbf{\Hat{x}}), \quad \alpha_{pi} \in \mathbb{R}^{+}, 
    \label{eq:totPILoss}
\end{equation}
where $\alpha_{pi}$ is a hyperparameter.
The PI-LSTM has shown accurate performance   in the reconstruction and forecasting of missing observations and successfully infers stability properties for the Lorenz-96 and Kuramoto-Sivashinsky, even in the presence of noise \citep{ozalp2023physics}.
The code of the PI-LSTM is available on Github \href{https://github.com/MagriLab/POLA-LSTM}{PI-LSTM}.

\subsection{Data Preparation}
LSTMs are employed for sequential data, wherein the order itself encapsulates temporal information, in contrast to explicitly incorporating time as a feature, as seen in feedforward neural networks. When employing LSTMs and backpropagation, additional steps become necessary. The data preparation for the LSTM consists of three steps. \\

\textbf{Normalization and temporal spacing of the input.}
Whilst LSTM require equidistantly time-spaced input, determining the optimal temporal spacing for the LSTM is not straightforward and depends on the characteristics of the dynamical system. The performance of the network  varies based on the temporal spacing selected. In instances where the temporal spacing between two observations is too small, the network may converge to a fixed point in the closed-loop prediction.  This occurs when temporal information available within a narrow time window is insufficient, hindering the LSTM's capacity to effectively capture the temporal dynamics of the system. Conversely, if the intervals are overly large, the predictive behaviour of the network tends to exhibit divergence, as the network struggles to interpolate between sparse data points.
It is advisable to treat the temporal sampling as a hyperparameter, and a recommended practice involves normalising the input data based on the activation function. Typically, this means normalizing to the range of $\begin{bmatrix}-1, 1\end{bmatrix}$. This normalisation contributes to stabilising the LSTM network's performance. \\ 
 
\textbf{Data splitting.} This step splits the dataset into training, validation and test sets. The training set is used to train the model parameters, the validation set helps fine-tune and optimize the model, and the test set assesses the model's performance on unseen data, providing an estimate of its generalization ability.
\begin{figure}
   \centering
    \includegraphics[width=\textwidth]{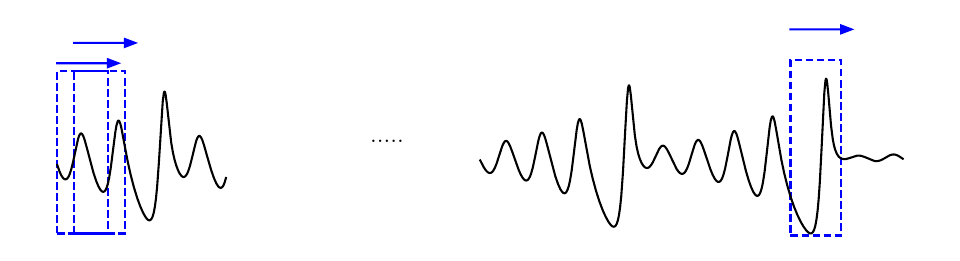}
    \caption{Illustration of the sliding window technique applied to a trajectory. A fixed time window is selected from the training data (indicated by the blue dashed line). The next time window is selected by shifting it along the temporal axis to process consecutive time windows. }
    \label{fig:lstm_timewindow}
\end{figure}
It is necessary to split the training data, originally presented as a long trajectory, into smaller time windows, as the LSTM training employs backpropagation through time. The sliding window approach involves selecting a fixed-size window from the training and subsequently shifting it along the temporal axis to process consecutive time windows, see Figure~\ref{fig:lstm_timewindow}. Careful consideration should be given to this hyperparameter of the window size, to ensure a balance between learning the short-term and long-term temporal dependencies. \\ 

\textbf{Window size selection. } 
The LSTM takes data of the following input shape (batch size, window size, dimension of observations). 
The window size refers to the number of consecutive time steps and should be chosen in consideration of the hidden state dimension. While increasing the window size directly impacts the number of backpropagation through time steps and training time, an alternative strategy involves increasing the hidden state dimension.  The memory capabilities of the LSTM are intricately linked to both the chosen window size and the dimension of the hidden state, making their calibration essential for the effective model performance.

\subsection{Jacobian of the LSTM}

Can LSTM infer the stability properties of solutions? The answer, as we know from Sec.~\ref{eq:tangent_eq}, lies in the Jacobian and its spectral properties. 
To compute the Lyapunov exponents as outlined in Sec.~\ref{subsec:algo}, the Jacobian of the system is required. The LSTM, when employed in closed-loop, defines a dynamical system.  
The Jacobian is the gradient of the internal states at a single timestep 
\begin{align}\label{coupled_Jac}
    \mathbf{J}_{LSTM}( \mathbf{c}_{i},  \mathbf{h}_{i}) = \begin{bmatrix} \dfrac{\partial \mathbf{c}_{i}}{\partial \mathbf{c}_{i-1}}  &\dfrac{\partial \mathbf{c}_{i}}{\partial \mathbf{h}_{i-1}} \\[2ex]
    \dfrac{\partial \mathbf{h}_{i}}{\partial \mathbf{c}_{i-1}} & \dfrac{\partial \mathbf{h}_{i}}{\partial \mathbf{h}_{i-1}} \end{bmatrix}, 
\end{align}
which analytically provided by~\citet{ozalp2023physics}
\begin{align}
\begin{split}
    \frac{\partial \mathbf{c}_{i}}{\partial \mathbf{c}_{i-1}} &= \mathbf{I}  *\mathbf{f}_{i},\\
\frac{\partial \mathbf{c}_{i}}{\partial \mathbf{h}_{i-1}} &=
    \mathbf{c}_{i} *\mathbf{f}_{i}* ( \mathbf{I}-\mathbf{f}_{i}) \mathbf{W}^f + \mathbf{i}_{i} *( \mathbf{I}-\mathbf{i}_{i}) \mathbf{W}^i *\Hat{\mathbf{c}}_i +\mathbf{i}_i * \left(\mathbf{I}-\mathbf{\Hat{C}}_{i}^2\right) \mathbf{W}^{g},\\
\frac{\partial \mathbf{h}_{i}}{\partial \mathbf{c}_{i-1}} &= \left( \mathbf{I} -            \tanh^2{\left(\mathbf{c}_{i} \right)} \right)*\mathbf{o}_{i} * f_i,\\
\frac{\partial \mathbf{h}_{i}}{\partial \mathbf{h}_{i-1}} &=  
    \mathbf{o}_{i}*( \mathbf{I}-\mathbf{o}_{i})*\tanh{(\mathbf{c}_{i})} + 
    \left(  \mathbf{I} - \tanh^2{(\mathbf{c}_{i})}\right)*\mathbf{o}_{i} * \\
    &\left(\mathbf{c}_{i} *\mathbf{f}_{i} *( \mathbf{I}-\mathbf{f}_{i}) \mathbf{W}^f + \mathbf{i}_{i} *( \mathbf{I}-\mathbf{i}_{i}) \mathbf{W}^i *\Hat{\mathbf{c}}_i +\mathbf{i}_i * \left( \mathbf{I}-\mathbf{\Hat{C}}_{i}^2\right) \mathbf{W}^{g}\right).
    \end{split}
    \label{eq:LSTM_Jac}
\end{align} 

%
%
\section{Tutorial: Lorenz system}
The Lorenz system \eqref{eq:lorenz63} is a deterministic nonlinear ordinary differential system, with three positive parameters $\sigma, \rho, \beta$ 
\begin{align}
		\dfrac{\mathrm{d}x}{\mathrm{d}t} = \sigma (y-x), \;\;\;
	\label{eq:lorenz63}
		\dfrac{\mathrm{d}y}{\mathrm{d}t} = x (\rho-z) - y, \;\;\;
		\dfrac{\mathrm{d}z}{\mathrm{d}t} = xy - \beta z.
\end{align}

For certain values of these parameters, the most common are $\sigma=10$, $\beta=8/3$ and $\rho=28$, for which the system has chaotic solutions\footnote{Note that the parameter $\rho$ in the Lorenz equations has nothing to do with the spectral radius in ESNs. This is a slight abuse of notation.}~\citep{Lorenz1963}. Given the  simplicity of the equations, low dimensionality of the system and the vast amount of research available on it, the Lorenz system is a prime candidate for playing around with chaos, and testing machine learning. 
First, we perform fixed-point analysis. 
Denoting the state vector ${\mathbf{x}}=(x,y,z)^\mathrm{T}$, the system can be rewritten as $\dot{\mathbf{x}} = \mathbf{F}({\mathbf{x}})$. The fixed points of the system are determined by solving $\dot {\mathbf{x}}=\mathbf{0 }\rightarrow \mathbf{F}({\mathbf{x}}) = \mathbf{0}$, which, apart from the trivial fixed point ${\mathbf{x}}^*=\mathbf{0}$, yields 

\begin{equation}
	\left\{
	\begin{matrix}
		x^* = y^* &=& \pm \sqrt{\beta(\rho-1)} \\
		z^* &=& \rho - 1
	\end{matrix}
	\right.	
\end{equation}

Therefore, for $\rho \leq 1$, only the trivial fixed point ${\mathbf{x}}^*=\mathbf{0}$ exists. At $\rho=1$, the system undergoes a pitchfork bifurcation, after which two new families of fixed points appear $\mathcal{C}^-$, $\mathcal{C}^+$. The linear stability of these can be determined by analysing the Jacobian matrix of the system~\eqref{eq:lorenz:jacobian} 

\begin{equation}
	\label{eq:lorenz:jacobian}
	{\mathbf{J}} \equiv \left. \frac{\mathrm{d}\mathbf{F}}{\mathrm{d}{\mathbf{x}}} \right |_{{\mathbf{x}}={\mathbf{x}}^*} = 
	\begin{pmatrix}
		-\sigma & \sigma & 0\\ 
		\rho - z^* & -1 & -x^*\\ 
		y^* & x^* & -\beta
	\end{pmatrix}
\end{equation}
For the trivial fixed point, ${\mathbf{x}}^*=0$, the equation in $z$ becomes decoupled and the eigenvalues are easily determined:

$$\lambda_1 = -\beta, \quad \lambda_{2,3} = \frac{-(\sigma+1) \pm \sqrt{(\sigma+1)^2+4\sigma (\rho-1)}}{2}$$

The origin is therefore linearly stable for $\rho < 1$ and linearly unstable for $\rho > 1$. For $\mathcal{C}^\pm$, the characteristic polynomial of $\mathbf{J}$ is a third-degree polynomial given in~\eqref{eq:lorenz:char_poly}.
\begin{equation}
	\label{eq:lorenz:char_poly}
	p(\lambda) \equiv \det(\mathbf{J}-\xi \mathbf{I}) = - \xi^3 - \xi^2 (\beta + \sigma + 1) - \xi \beta (\sigma+\rho) - 2 \beta \sigma (\rho-1)
\end{equation}

Instead of trying to solve a third-degree polynomial equation $p(\xi)=0$ directly, one can look for a Hopf bifurcation, where a complex conjugate pair crosses the imaginary axis, by setting $\xi = i \mu$ with $\mu \in \mathbb{R}$. This results in two equations: one for the real and one for the imaginary part. By solving each one for $\mu$ and equating the results, the stability condition~\eqref{eq:lorenz:stab_cond} is obtained.

\begin{equation}
	\label{eq:lorenz:stab_cond}
	\rho < \sigma \frac{\sigma + \beta + 3}{\sigma - \beta - 1}
\end{equation}

\begin{figure}[!ht]
    \centering
    \includegraphics[width=\textwidth]{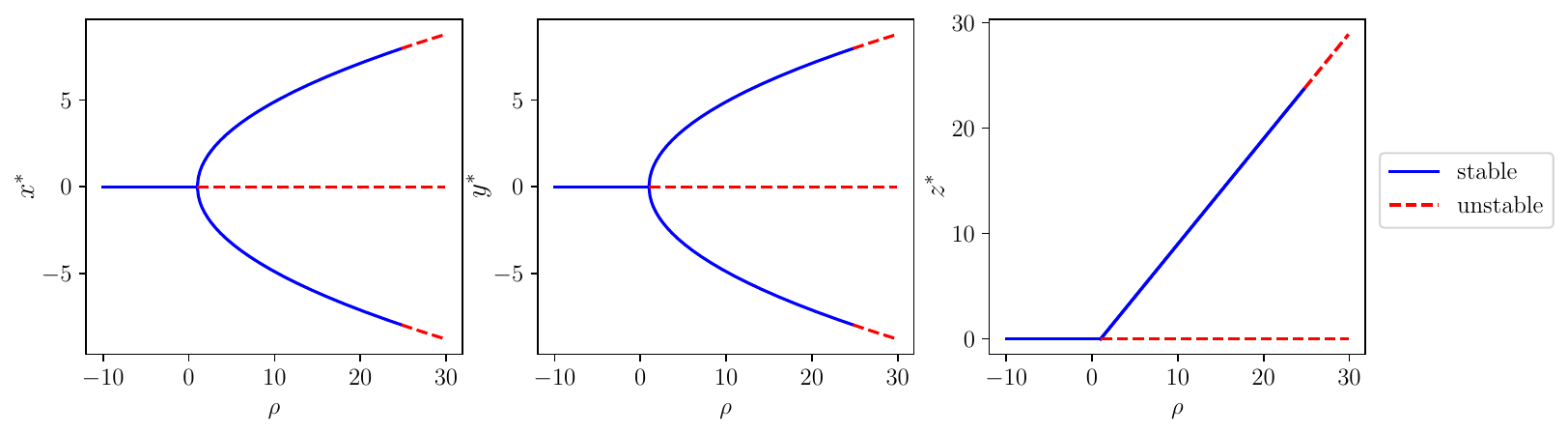}
    \caption{Bifurcation diagram for $\beta=8/3, \sigma=10$.}
    \label{fig:bifurcation_lorenz63}
\end{figure}

With the classic parameter values $\sigma=10$, $\beta=8/3$ and $\rho=28$, all three fixed points are unstable. A bifurcation diagram can be found in Figure~\ref{fig:bifurcation_lorenz63}.

%
%
%

\begin{figure}[!ht]
    \centering
 %
\includegraphics[width=\textwidth]{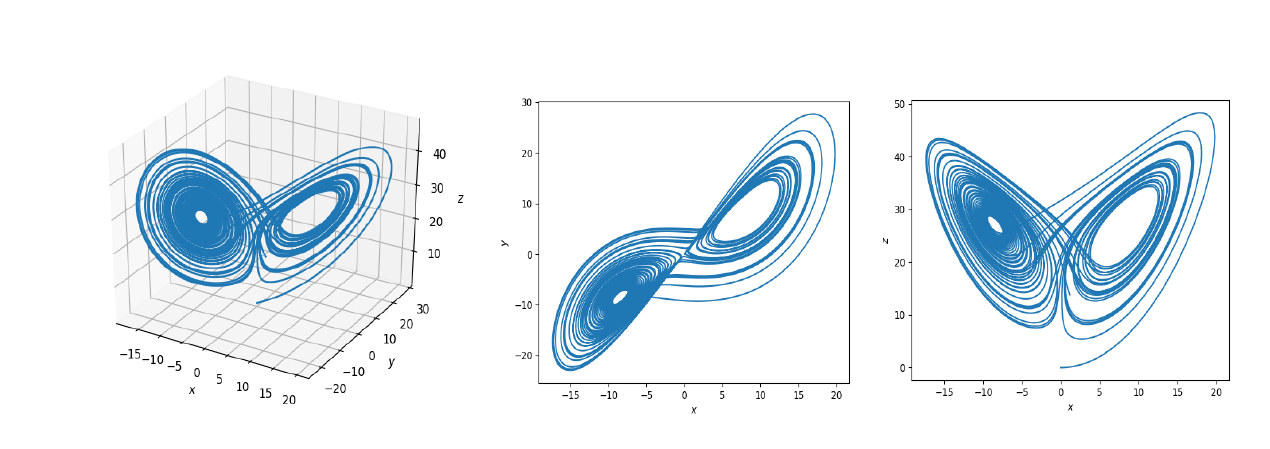}
    \caption{Strange attractor in the Lorenz system.}
    \label{fig:lorenz:butterfly}
\end{figure}

Second, we explore the chaotic regime.
With the classic parameter values $\sigma=10$, $\beta=8/3$ and $\rho=28$, the trajectories will converge towards the Lorenz attractor. Figures~\ref{fig:lorenz:butterfly} show three-dimensional, $x$-$y$ and $x$-$z$ views of the trajectory with initial condition ${\mathbf{x}}_0 = (10^{-9}, 10^{-9}, 10^{-9})^\mathrm{T}$ up to $t=50$.

Third, we characterize the chaotic regime. 
Figure~\ref{fig:lorenz:lyapunov:growth} is a graph of the exponential growth of the separation trajectory, $\Delta {\mathbf{x}}$, between the trajectory starting at ${\mathbf{x}}_0 = (-8.67,4.98,25.00)^\mathrm{T}$ and at ${\mathbf{x}}_0+\Delta {\mathbf{x}}_0$, with $\Delta {\mathbf{x}}_0 = (0, 0, 10^{-9})^\mathrm{T}$. It shows that $\log ||\frac{\Delta {\mathbf{x}}(t)}{\Delta {\mathbf{x}}_0}||$ starts at $0$, follows a linear growth until $t=25$, finally reaching a plateau. The largest Lyapunov exponent is calculated by a linear regression applied in $t \in [0, 25]$ and its value is $\lambda_1 = 0.929$. To obtain a better estimate, the calculated value of $\lambda_1$ should be averaged over many simulations.

\begin{tcolorbox}[breakable, opacityframe=.1, title=Tutorial: \href{https://github.com/MagriLab/Tutorials/tree/main}{MagriLab/Tutorials}
 ]

We present a tutorial on the application of LSTM and ESN architectures for modelling the Lorenz 63 system. The Lorenz data is generated using an explicit 4th-order Runge-Kutta method with $\Delta t =0.01$.
For training purposes, we use $100 t_p$, and both validation and testing are also implemented. The ESN implementation involves a hyperparameter sweep, considering the network's high sensitivity to parameters. Notably, the LSTM requires a longer training time but demonstrates greater robustness to parameter changes, due to its utilisation of backpropagation. The LSTM results below are presented with a hidden dimension $30$. In comparison, the reservoir of the ESN has to be at least of dimension $100$ to achieve comparative results. 
All results and code below can be found on \href{https://github.com/MagriLab/Tutorials/tree/main}{Github}. 

\end{tcolorbox}

 \begin{tcolorbox}[breakable, opacityframe=.1, title=Algorithm: Closed-loop prediction.
 ]
\noindent \textbf{Option 1 - LSTM:}\\
\textbf{Input:} Observations $\mathbf{\Hat{x}}(t_{i+1})$ for the window size\\
\begin{enumerate}
  \item \textbf{Open-loop. } $\text{for }i=0, \dots, N_{window}:$
  \begin{align*}
  \mathbf{\Hat{x}}(t_{i+1}),& \mathbf{c}_{i+1},\mathbf{ h}_{i+1} = LSTM(\mathbf{x}(t_{i}), \mathbf{c}_{i}, \mathbf{h}_{i})
  \end{align*}
 \item \textbf{Closed-loop. } $\text{for }i=N_{window}, \dots, N_{prediction}:$
   \begin{align*}
   \mathbf{\Hat{x}}(t_{i+1}),& \mathbf{c}_{i+1},\mathbf{ h}_{i+1} = LSTM(\mathbf{\Hat{x}}(t_{i}), \mathbf{c}_{i}, \mathbf{h}_{i})  \end{align*}
\end{enumerate}

\noindent \textbf{Option 2 - ESN:}\\
\textbf{Input:} Observations $\mathbf{\Hat{x}}(t_{i+1})$ for the washout size\\
\begin{enumerate}
  \item \textbf{Open-loop. } $\text{for }i=0, \dots, N_{washout}:$
  \begin{align*}
  \mathbf{\Hat{x}}(t_{i+1}), \mathbf{r}(t_{i+1}) = ESN(\mathbf{x}(t_{i}), \mathbf{r}(t_{i+1}))
  \end{align*}
 \item \textbf{Closed-loop. } $\text{for }i=N_{washout}, \dots, N_{prediction}:$
   \begin{align*}
\mathbf{\Hat{x}}(t_{i+1}), \mathbf{r}(t_{i+1}) = ESN(\mathbf{\Hat{x}}(t_{i}), \mathbf{r}(t_{i+1}))\end{align*}
\end{enumerate}

\end{tcolorbox}

\begin{figure}[!ht]
    \centering
    \includegraphics[width=0.9\textwidth]{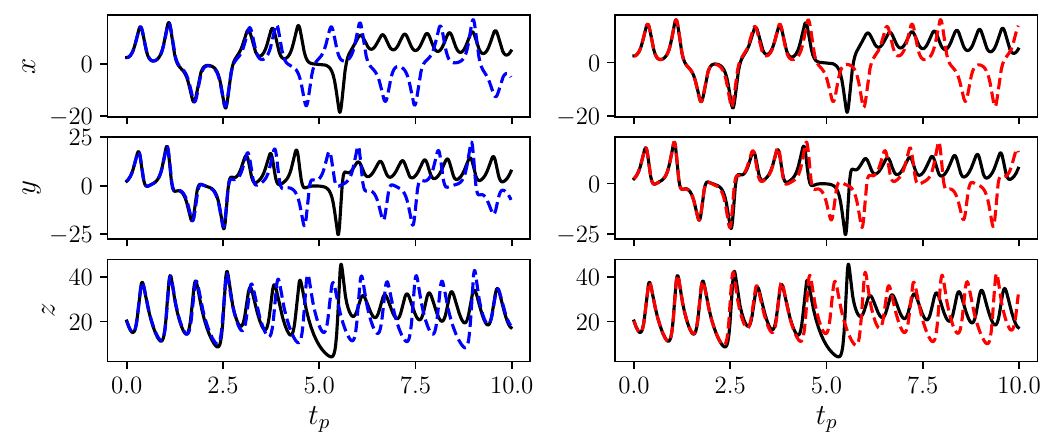}
    \caption{Closed-loop prediction of LSTM (blue dashed line) and ESN (red dashed line) compared to the test data (black line).}
    \label{fig:closed-loop_predictions}
\end{figure}

\noindent\textbf{Short-term prediction of the ESN and LSTM}\\
\noindent For the evaluation of the trained ESN and LSTM, both networks are employed in closed-loop mode on unseen test data.  In Figure~\ref{fig:closed-loop_predictions}, the networks are compared based on their short-term prediction capabilities. Due to the inherent chaotic nature of the data, we anticipate a divergence over time. Initially, both models closely track the trajectory for approximately $4t_p$; however, thereafter, the trajectories start to diverge. 

\noindent

   
    


    \label{tab:avg_ph_esn_lstm}
Following the metrics in Section~\ref{sec:validation}, we can compute the prediction horizon of both models on the test data. Based on the closed-loop prediction, we evaluate the NRMSE from Eq.~\eqref{eq:NRMSE} and the PH from Eq.~\eqref{eq:PH} with $k_{PH} = 0.4$. Due to the inherent chaotic nature of the data, which leads to varying prediction horizons depending on the test interval, Single Shot testing becomes unreliable. In the table below, we provide a comparison of prediction horizons averaged over $M$ test intervals to address this variability in Table~\ref{tab:avg_ph_esn_lstm}.
\begin{table}[h]
 \centering
\begin{tabular}{ |p{3cm}||p{3cm}|p{3cm}|   }
 \hline
$M$& ESN & LSTM \\
 \hline
  \hline
 1 & $ 3.16 t_p$ & $ 4.41 t_p$ \\
 \hline
10 & $ 5.27 t_p$ & $4.24 t_p$ \\
 \hline
 100 & $ 5.53 t_p$ & $ 3.96 t_p$ \\
 \hline
\end{tabular}  
\caption{Average prediction horizon over $M$ intervals with $k_{PH}=0.4$.}
    \label{tab:avg_ph_esn_lstm}
\end{table}

\begin{figure}[!ht]
    \centering
    \includegraphics[width=0.9\textwidth]{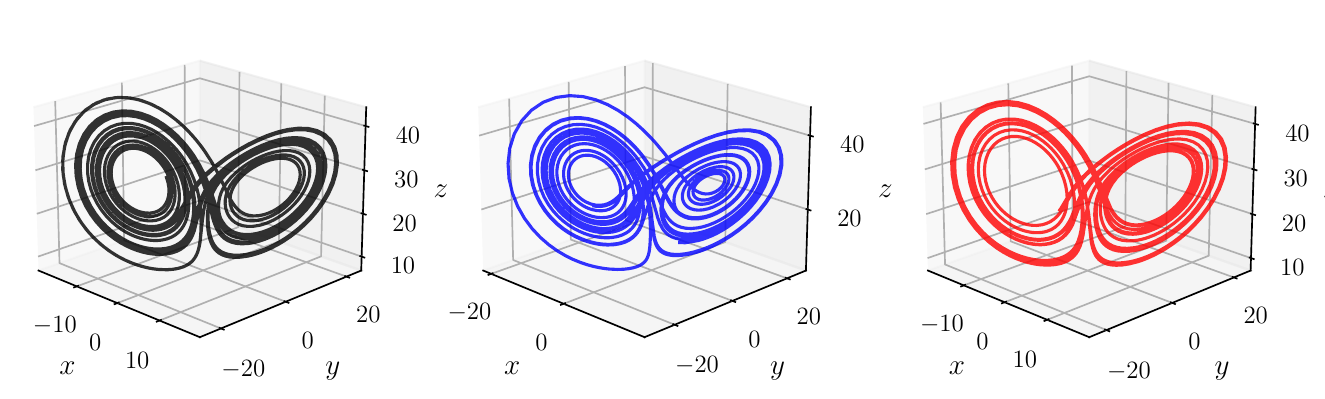}
    \caption{Attractor of the test data (left), LSTM (middle) and ESN (right).}
    \label{fig:attractor_reconstruction}
\end{figure}

For a visual inspection, we plot the prediction of the networks in the phase space, see Figure~\ref{fig:attractor_reconstruction}. Both the LSTM and ESN attractors exhibit the characteristic butterfly shape, indicating that they recover the dynamical properties of the system.  

\begin{figure}
    \centering
   \includegraphics[width=0.9\textwidth]{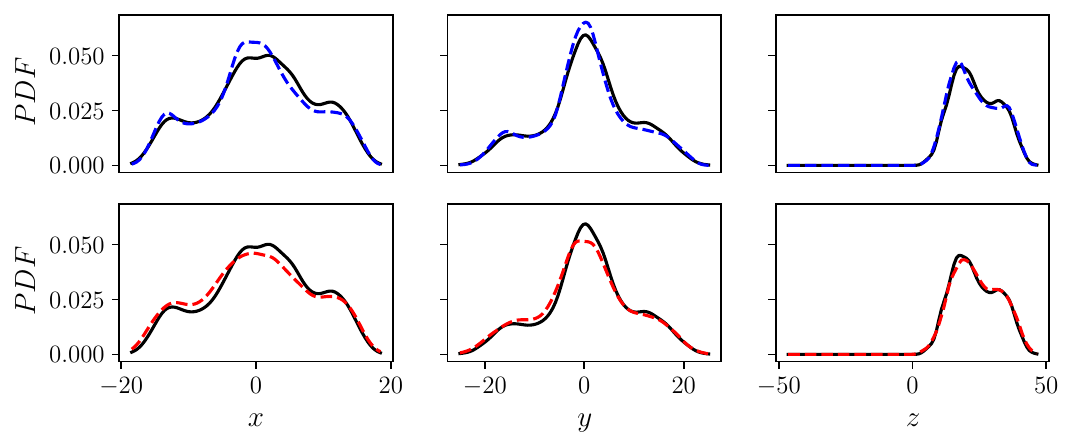}
    \caption{Probability density function (PDF) of the closed-loop prediction of LSTM (blue dashed line) and ESN (red dashed line) compared to the test data (black line) for $500t_p$.}
    \label{fig:stastics}
\end{figure}

To assess the networks' performances, another metric involves tracking the prediction statistics through a probability density function (PDF) estimate, as illustrated in Figure~\ref{fig:stastics}. 

A more effective performance assessment involves analysing the Lyapunov spectrum of the networks. The Lyapunov exponents serve as indicators of the network's accuracy in inferring stability properties and its ability to reproduce the chaotic nature in the prediction. Both networks, the ESN and LSTM, are employed for the stability analysis. In Table~\ref{tab:le_esn_lstm}, the inferred Lyapunov exponents are presented. Both networks infer the spectrum by reproducing a positive, neutral and negative exponent. Both networks can also be employed to analyse further stability properties, such as covariant Lyapunov vectors \citep{oezalp_chaos_2023, margazoglou2023stability}.
 \begin{tcolorbox}[breakable, opacityframe=.1, title=Algorithm: Computing Lyapunov spectrum with the LSTM/ESN.
 ]
\textbf{Initialisation:  Repeat Steps 1.-3. from Section \ref{sec:practical_computation_le}.}\\
\newline
\noindent \textbf{Evolve the solution and GSV simultaneously for $N^{lyap}$ steps.}

\noindent \textbf{Option 1 - LSTM:}
\begin{enumerate}
  \item {Evolve the system with the LSTM} $ \mathbf{\Hat{x}}(t_{i+1}), \mathbf{c}_{i+1},\mathbf{ h}_{i+1} = LSTM(\mathbf{\Hat{x}}(t_{i}), \mathbf{c}_{i}, \mathbf{h}_{i}) $
     \item {Compute the LSTM Jacobian: } $\mathbf{J} \gets Jac_{LSTM}(\mathbf{c}_{i+1}, \mathbf{h}_{i+1})$
\end{enumerate}

\noindent \textbf{Option 2 - ESN:}
\begin{enumerate}
  \item {Evolve the system with the ESN} $ \mathbf{\Hat{x}}(t_{i+1}), \mathbf{r}(t_{i+1}) = ESN(\mathbf{\Hat{x}}(t_{i}), \mathbf{r}(t_{i+1}))$
     \item {Compute the ESN Jacobian: } $\mathbf{J} \gets Jac_{ESN}(\mathbf{r}(t_{i+1}))$
\end{enumerate}

\textbf{ Discard a transient.}
\begin{enumerate}
 \setcounter{enumi}{2}
     \item {Update linearized solution:} $\mathbf{U} \gets \mathbf{J}\mathbf{U}$
     \item {Orthonormalize and update Gram Schmidt vectors:} $\mathbf{Q}, \mathbf{R} \gets QR(\mathbf{U}); \mathbf{U} \gets \mathbf{Q}$
     \item {Track Lyapunov exponents:} $\mathbf{\lambda}[:, i] \gets \mathrm{log}(diag(\mathbf{R}))/\Delta t$ 
\end{enumerate}
\textbf{Time-averaged Lyapunov exponents: } $ \mathbf{\lambda}_j \gets \sum_{i=0}^{N_{QR}}  \mathbf{\lambda}[j, i]/T_{lyap}$ 
\end{tcolorbox}

\begin{table}[h]
 \centering
\begin{tabular}{ |p{3cm}||p{3cm}|p{3cm}|p{3cm}|    }
 \hline
 \multicolumn{4}{|c|}{Lorenz 63} \\
 \hline
$\lambda_i$&  target & ESN & LSTM \\
 \hline
 $1$ & $0.9050$ & $0.9067$ & $0.873$ \\
 $2$ &   $9 \times 10^{-5}$& $-8 \times 10^{-5}$& $-8 \times 10^{-3}$\\
 $3$ & $-14.572$  & $-14.664$ & $-14.0959$\\
 \hline
\end{tabular}   
\caption{Lyapunov Exponents for LSTM and ESN.}
\label{tab:le_esn_lstm}
\end{table}

\clearpage

\section{Ridge regression for ESN training}\label{app:RidgeReg}

The weights of the output matrix $\mathbf{W}_{out}$ are obtained by solving 
\begin{align}\label{eq:mini}
    \begin{matrix}
        \mathrm{arg min}\\
        {\mathbf{W}_{out}}
    \end{matrix}&\quad\dfrac{1}{N}{\sum_{i=1}^{N}\left\|\mathbf{W}_{out}\mathbf{r}(t_i) - \mathbf{x}(t_i)\right\|_2^2 + \frac{\gamma}{N_x}\sum_{j=1}^{N_x} \left\|\mathbf{w}_{{out,j}}\right\|_2^2} = \mathcal{J}(\mathbf{W}_{out}) \\[1em] \nonumber
    \mathrm{s.t.} &\quad \mathbf{r}(t_i)= \mathrm{tanh}\left(\sigma_\mathrm{in}\mathbf{W}_{in}\mathbf{x}(t_{i-1})+
        \rho\mathbf{W}\mathbf{r}(t_{i-1})\right),\\\nonumber
    &\quad\mathbf{r}_0=\mathbf{0},
\end{align}
where $\gamma$ is the Tikhonov regularisation parameter, 
$\sigma_\mathrm{in}$ is the input scaling factor, and
$\rho$ is the spectral radius. 
  The minimisation problem \eqref{eq:mini} has an analytical solution, which is obtained by minimizing the cost function $\mathcal{J}$
with respect to the output matrix $\mathbf{W}_{out}$, and setting the result to zero, such that
\begin{align}
    \nonumber
    \dfrac{\mathrm{d}\mathcal{J}}{\mathrm{d}\mathbf{W}_{out}} &=  \dfrac{1}{NN_x}\sum_{i=1}^{N}\left\{2\left(\mathbf{W}_{out}\mathbf{r}(t_i) - \mathbf{x}(t_i)\right)\mathbf{r}(t_i)^\mathrm{T} + 2\gamma \mathbf{W}_{{out}}\right\} \\\nonumber
     \label{eq:dJ_ESN}
     &=  \dfrac{1}{NN_x}\sum_{i=1}^{N}2\left\{\left(\mathbf{W}_{out}\mathbf{r}(t_i)\mathbf{r}(t_i)^\mathrm{T} + \gamma \mathbf{W}_{out}\right) - \mathbf{x}(t_i)\mathbf{r}(t_i)^\mathrm{T}\right\} = \mathbf{0}.
\end{align}
Rearranging the terms in \eqref{eq:dJ_ESN} we find that
\begin{align}
 \sum_{i=1}^{N}\mathbf{W}_{out}\left(\mathbf{r}(t_i)\mathbf{r}(t_i)^\mathrm{T} + \gamma \mathbf{I}\right) = \sum_{i=1}^{N}\mathbf{x}(t_i)\mathbf{r}(t_i)^\mathrm{T} \quad\Rightarrow
 \quad \sum_{i=1}^{N}\left(\mathbf{r}(t_i)^\mathrm{T}\mathbf{r}(t_i) + \gamma \mathbf{I}\right)\mathbf{W}_{out}^\mathrm{T} = \sum_{i=1}^{N}\mathbf{r}(t_i)\mathbf{x}(t_i)^\mathrm{T},
\end{align}
which can be written in the compact form
\begin{equation}
    \left(\mathbf{R}\mathbf{R}^\mathrm{T} + \gamma \mathbf{I}\right)\mathbf{W}_{out}^\mathrm{T} = \mathbf{R} \mathbf{X}^\mathrm{T},
\end{equation}
where 
$\mathbf{R} = \left[\mathbf{r}(t_{1})\,|\dots|\,\mathbf{r}(t_{N})\right]$  
and ${\mathbf{X}}=\left[\mathbf{x}(t_1)\,|\dots|\,\mathbf{x}(t_{N})\right]$ are the horizontal time-concatenation of the output augmented reservoir state and training data. 
The hyperparameters $\sigma_\mathrm{in}, \rho$ and $\gamma$ can be optimized during training through Recycle Validation \citep{racca2021robust}.

\medskip

\bibliographystyle{apalike}
\bibliography{references}
\end{document}